\documentclass[a4paper,11pt]{article}

\usepackage{physics}
\usepackage[usenames,dvipsnames]{xcolor}
\usepackage{graphicx}
\usepackage{slashed}
\usepackage{soul}
\usepackage[normalem]{ulem}
\pdfoutput=1 

\usepackage{jheppub} 

\hypersetup{%
  colorlinks=true, linktocpage=true, pdfstartpage=1, pdfstartview=FitV,
  breaklinks=true, pageanchor=true,
  pdfpagemode=UseNone,
  plainpages=false, bookmarksnumbered, bookmarksopen=true, bookmarksopenlevel=1,
  hypertexnames=true, pdfhighlight=/O,
  urlcolor=RoyalBlue, linkcolor=ForestGreen, citecolor=RoyalBlue,
  pdftitle={Scalar leptoquark couplings at low and high energies},
  pdfauthor={Innes Bigaran, Rodolfo Capdevilla, Ulrich Nierste},
  pdfsubject={},
  pdfkeywords={},
  pdfcreator={pdfLaTeX},
  pdfproducer={LaTeX with hyperref}
}

\preprint{{\small FERMILAB-PUB-23-309-T, TTP24-030, P3H-24-055}}

\newcommand{\no}{\nonumber}
\newcommand{\nn}{\nonumber \\}
\newcommand{\lt}{\left}
\newcommand{\rt}{\right}
\renewcommand{\real}{\mbox{Re}\,}

\newcommand{\tev}{\,\mbox{TeV}}

\newcommand{\fig}[1]{Fig.~\ref{#1}}

\newcommand{\eq}[1]{Eq.~(\ref{#1})}
\newcommand{\eqsand}[2]{Eqs.~(\ref{#1}) and (\ref{#2})}
\newcommand{\eqsto}[2]{Eqs.~(\ref{#1}--\ref{#2})}
\newcommand{\ov}{\overline}

\newcommand{\sls}{\!\!\! / \;} 

\author{Innes Bigaran$^{a,b}$,}
\author{Rodolfo Capdevilla$^a$, and}
\author{Ulrich Nierste$^{c}$}

\affiliation{$^a$ Theory Division, Fermi National Accelerator Laboratory, 
  Batavia, IL 60510-500, USA,\\
  $^b$ Northwestern University, Department of Physics \& Astronomy, 2145 Sheridan Road, Evanston, IL 60208, USA \\
  $^c$Institute for Theoretical Particle Physics, Karlsruhe Institute of Technology (KIT),\\
  Wolfgang-Gaede-Str. 1, 76131 Karlsruhe, Germany}

\emailAdd{ibigaran@fnal.gov}
\emailAdd{rcapdevi@fnal.gov}
\emailAdd{ulrich.nierste@kit.edu}

\abstract{
Scalar leptoquarks (LQ) with masses between 2 TeV and 50 TeV are prime candidates to explain deviations between measurements and Standard-Model predictions in decay observables of $b$-flavored hadrons (``flavor anomalies'').  
Explanations of low-energy data often involve ${\cal O}(1)$ LQ-quark-lepton Yukawa  couplings, especially when collider bounds enforce a large LQ mass. This calls for the calculation of radiative corrections involving these couplings.
Studying such corrections to LQ-mediated $b\to c\tau \nu$ and $b\to s\ell^+\ell^-$ 
amplitudes, we find that they can be absorbed into finite renormalizations of the 
LQ Yukawa couplings. If one wants to use Yukawa  couplings extracted from low-energy data 
for the prediction of on-shell LQ decay rates, one must convert the low-energy couplings to their high-energy counterparts, which subsume the corrections to the on-shell LQ-quark-lepton
vertex. We present compact formulae for these correction factors and find that 
in scenarios with $S_1$, $R_2$, or $S_3$ LQ the high-energy coupling is always smaller than the low-energy one, which weakens the impact of collider data on the determination of the allowed parameter spaces. For the $R_2$ scenario addressing $b\to c\tau \nu$, in which one of the two involved Yukawa coupling  must be significantly larger than 1, we find this coupling reduced by 15\% at high energy. 
If both $S_1$ and $R_2$ are present, the high-energy coupling can also be larger and 
the size of the correction is unbounded, because tree contribution and vertex corrections involve different couplings. We further present the conversion formula to the 
$\overline{\rm MS}$  scheme for the Yukawa couplings of the $S_3$ scenario.}

\title{Radiative corrections relating leptoquark-fermion couplings probed at low and high energy}

\begin{document}

\maketitle
\flushbottom

\newpage
\setcounter{page}{1}
\section{Introduction}
\label{sec:intro}

\begin{figure}[t!]
 \includegraphics[height=3.5cm,clip=true]{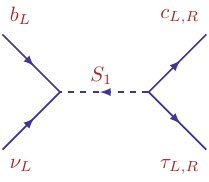}\hfill
\includegraphics[height=3.5cm,clip=true]{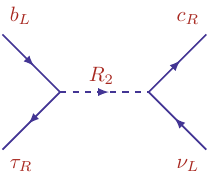}\hfill
 \includegraphics[height=3.5cm,clip=true]{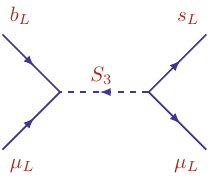}
  \caption{Contributions of scalar leptoquarks to the decay modes
    $b\to c \tau^- \bar\nu$ and $b\to s \mu^+\mu^-$.\label{fig:bdec}}
  ~\\[-4mm]
  \hrule
\end{figure}

Leptoquarks~(LQs) are hypothetical particles that directly couple to Standard Model~(SM) leptons and quarks. They feature in many models for Beyond the Standard Model~(BSM) physics and, in particular, TeV-scale leptoquarks remain phenomenologically well-motivated candidates 
{(see e.g. Refs.~\cite{Raj:2016aky,Diaz:2017lit,Bansal:2018eha,Schmaltz:2018nls,Bhaskar:2021gsy,Bernigaud:2021fwn,Dorsner:2022ibm,Goncalves:2023qpz,Bhaskar:2023ftn})}
for searches at the Large Hadron Collider driven by the quark transitions $b\to s \ell^+\ell^-$, where $\ell=e,\mu$,~\cite{LHCb:2015ycz,LHCb:2015wdu,ATLAS:2018cur,CMS:2018qih, LHCb:2018jna, CMS:2019bbr,BELLE:2019xld,CMS:2020oqb,LHCb:2020dof,LHCb:2020gog, LHCb:2021vsc,LHCb:2021trn,LHCb:2021xxq,LHCb:2021lvy,LHCb:2021zwz, LHCb:2022qnv,LHCb:2022zom} and $b\to c \tau \nu$~\cite{HFLAV:2022esi}\footnote{HFLAV~\cite{HFLAV:2022esi} combines experimental results of Refs.~\cite{BaBar:2012obs,BaBar:2013mob,Belle:2019rba,LHCb:2023cjr,Belle:2015qfa,Belle:2016dyj,Belle:2017ilt} with
the theory predictions of Refs.~\cite{Bigi:2016mdz,Gambino:2019sif,Bordone:2019vic,Bernlochner:2017jka,Jaiswal:2017rve,BaBar:2019vpl,Martinelli:2021onb}.}, because several observables in these decay channels  deviate from their Standard-Model
(SM) predictions. To date, scalar leptoquarks (LQs) are the most popular particle species postulated to remedy
these \emph{flavor anomalies} (see e.g. Refs.~\cite{Sakaki:2013bfa,Dorsner:2016wpm,
  Dumont:2016xpj, Li:2016vvp,Bhattacharya:2016mcc,Chen:2017hir,
Crivellin:2017zlb,Cai:2017wry, Jung:2018lfu,Aydemir:2019ynb,Popov:2019tyc,Crivellin:2019dwb,Bigaran:2019bqv,Bansal:2018nwp,Iguro:2020keo,Ciuchini:2022wbq,Dev:2024tto}). The presence of these anomalies have been consolidated over the last decade and the scientific discussion in the theory community addresses the question whether their origin is BSM physics or incorrectly calculated SM predictions. 

In the case of the ratios of branching fractions, $R(D^{(*)})=\text{BR}(B\to D^{(*)} \tau \nu)/\text{BR}(B\to D^{(*)} \ell \nu)$ ($\ell=e,\mu$),
the theory prediction is very robust, the only non-perturbative quantity entering the prediction is a ratio of form factors multiplying a term suppressed by the mass ratio $m_\tau^2/m_B^2$. Changing this form factor ratio to the level that the SM prediction for $R(D^{*})$ complies  with data leads to severe tensions in the predictions of measured polarisation data 
in $B\to D^* \ell\nu$ decays with light leptons $\ell=e,\mu$ and rule this explanation out \cite{Fedele:2023ewe}. A recent analysis employing maximal experimental information on form factor shapes found a deviation of the combined $b\to c\tau\nu$ data with the SM predictions at a level of 4.4$\sigma$ \cite{Iguro:2024hyk}.

For $b\to s \ell^+ \ell^-$ data, the significance of the anomaly is even higher. Since deviations with the same sign are observed in different hadronic decays, a SM explanation must invoke mistakes in several form factors or a correlated effect in all of these modes. So far only one explanation along the latter idea has been proposed, a radical enhancement of a charm loop contribution by a yet unknown non-perturbative effect. Such an effect would come with a different dependence of the decay amplitude on the invariant mass $q^2$ of the pair of leptons, but two studies of this dependence have found no evidence for a deviation compared to the theory prediction \cite{LHCb:2023gpo,Bordone:2024hui}. 
(Heavy BSM physics changes the overall size of the effect, but not the shape of the $q^2$-dependence.) Moreover, a theoretical calculation employing heavy-hadron chiral perturbation theory found a contribution from the charm loop which is far too small to explain the data \cite{Isidori:2024lng}, supporting similar findings from dispersive analyses \cite{Capdevila:2023yhq}. In light of these developments, an explanation of the $b\to s \ell^+ \ell^-$ data in terms of non-perturbative hadronic effects is disfavored, which in turn motivates 
a BSM explanation.

\fig{fig:bdec} illustrates the contributions 
of the scalar leptoquarks $S_1\sim (\bar 3, 1, 1/3)$, $R_2\sim( 3, 2,
7/6)$, and $S_3~\sim (\bar 3, 3, 1/3)$ to the
quark decay amplitudes of interest.\footnote{Here the numbers in brackets
indicate the quantum numbers with respect to the SM gauge group
$SU(3)_c \otimes SU(2)_L \otimes U(1)_Y$.} Additionally, for example, both $S_1$ and $R_2$ are capable of generating large chirally-enhanced contributions to the muon and electron anomalous magnetic moments (see e.g. Ref.~\cite{Bigaran:2020jil}), and all three leptoquarks may play a role in models for radiative neutrino mass generation (see e.g. Ref.~\cite{Gargalionis:2020xvt}).

 Typically, low-energy physics observables enable the determination of the ratio between the product of the couplings involved and the square of the LQ mass. This, in conjunction with current constraints on LQ searches, defines a {\it target parameter space} to be explored at high-energy colliders. To avoid present collider search bounds, LQ masses should exceed approximately 2 TeV (see e.g. Refs.~\cite{CMS:2018qqq,ATLAS:2019qpq}). Consequently, a model incorporating LQs such as $S_1$ or $R_2$ to explain the $b \to c \tau \nu$ data must possess $\mathcal{O}(1)$ LQ-quark-lepton couplings to generate a sufficiently large effect. For $S_3$, there is more flexibility -- $\mathcal{O}(1)$ couplings are only necessary for a LQ mass around 30 TeV \cite{Fedele:2023rxb}, a value far from being reached at the LHC. Therefore, at least for $S_1$ and $R_2$ significant radiative corrections are expected from loops involving these couplings, implying that the couplings affecting low-energy data may significantly differ numerically from those influencing high-energy phenomenology. \footnote{
 Corrections of this kind for vector LQs in a model with SU(4) gauge symmetry have been calculated in~\cite{Fuentes-Martin:2019ign,Fuentes-Martin:2020luw}.}
 
 Moreover, any particular model built for a given phenomenological purpose may contain multiple leptoquarks. For example they may come in several copies, e.g. at least two copies are needed for $S_3$ to address $b\to s \ell \ell$ data: $S_{3\,{e}}$ couples to $e$ but not to $\mu$ while for $S_{3\,{\mu}}$ the situation is opposite.\footnote{A single LQ $S_3$ coupling to both lepton species would lead to intolerably large $\mu\to e$ transitions~\cite{Fedele:2023rxb}.} Therefore, the interplay between the presence of multiple LQs for such aforementioned radiative corrections could provide new ways to indirectly search for evidence of multi-particlular models.

 Direct searches for new heavy particles at the LHC have thus far yielded no success, indicating a significant gap between the electroweak (EW) scale and the scale of BSM physics. Previous investigations into LQs in low-energy observables have incorporated QCD radiative corrections in the conventional manner, by matching the LQ-mediated amplitudes to the weak effective theory (WET) and employing renormalization group equations (RGEs) to evolve the resulting Wilson coefficients (WCs) from the renormalization scale $\mu_{\text{high}} \sim M_{\text{LQ}}$ (denoting the mass of the relevant LQ) to $\mu_{\text{low}} \sim m_b$, relevant to the $b$-flavored hadron decays of interest.\footnote{Fits referenced later in this work are done at $\mu_b=4.8$ GeV which is of the same scale as $m_b$.} This procedure effectively sums large logarithms multiplied by the strong coupling $\alpha_s$.\footnote{Electroweak running from the $M_{\text{LQ}}$ scale to the weak scale, necessitating an intermediate matching to operators of the Standard Model Effective Theory (SMEFT), is relatively minor compared to QCD running and is therefore neglected for the purposes of this study. Leptoquark production at hadron colliders at NLO in QCD has been extensively studied (see e.g. Refs.~\cite{Kramer:1997hh,Kramer:2004df,Dorsner:2018ynv,Mandal:2015lca,Borschensky:2020hot,Borschensky:2021hbo,Borschensky:2021jyk,Borschensky:2022xsa,Haisch:2022afh}). 
 }
 The focus of this paper, however, are genuine additional correction factors involving LQ couplings rather than $\alpha_s$, and solely enter the WCs at $\mu_{\rm high}$. In this way, these corrections modify the initial conditions for the WCs and the corrected coefficients can simply be evolved to $\mu_{\rm low} $ with the standard leading-log RGEs. 

When solely performing calculations of high-scale quantities, such as LQ production cross sections and LQ decay rates, all couplings come into play at the renormalization scale $\mu_{\text{high}}$, and no RGE running or matching procedure is required. Radiative one-loop corrections involving LQ-fermion couplings do not involve large logarithms, and in both high-energy and low-energy observables, these couplings are typically probed at the renormalization scale $\mu_{\text{high}}$. An important distinction between collider and flavor physics observables is the relevant momenta $p_i^{{\mu}}$ entering the loop diagrams: in collider observables, $|p_i^{{\mu}}| \approx M_{\text{LQ}}$, whereas in $b$ decays, $|p_i^{{\mu}}| \approx m_b \approx 0$. When we refer to ``energy" in this paper, we are referring to the magnitude of these momenta $p_i^{{\mu}}$, i.e., the energy scale at which the LQ interaction is probed, and not the renormalization scale $\mu$.

If one determines the low-energy LQ-fermion couplings from flavor observables, they can use the results of this paper to calculate the corresponding couplings at high energies. The tree amplitude then expressed in terms of these high-energy couplings contains the one-loop corrections to the LQ decay rate exactly. If one next wants to predict the corrections to the LQ production cross section at the LHC (or another high-energy collider), there will be additional loop diagrams to consider, which are, however, specific to the considered production process. 

For instance, a quark-gluon initiated process involves diagrams with off-shell LQ or fermion lines at the production vertex, each coming with different kinematics and, furthermore, a gluon-LQ-LQ vertex. The process-dependent corrections can be consistently calculated in the renormalization scheme in which the  LQ-quark-lepton coupling is defined at the high-energy scale, which breaks the radiative corrections up into a piece contained in  our calculation and the process-specific remainder. Thus our corrections do not fully subsume all radiative corrections to LQ production.

The corrections calculated in this work are universal, renormalization-scheme independent, and capture all effects related to the LQ decay vertex. These may therefore be easily included in simulations with tree-level event generators. 

Note that irrespective of whether the flavour anomalies persists or not, any future study of collider searches for LQs must necessarily take the constraints from low-energy data into account. Therefore the relevance of the corrections presented in this paper does not depend on the fate of the anomalies. On the contrary, if future data on $b\to s \ell^+\ell^-$ and $b\to c\tau\nu$ exactly comply with the SM, the constraints on the LQ couplings and masses derived from these data will be strongest as there will be  no leeway for BSM contributions.

In Section \ref{sec:setup} we explain our approach by providing a definition of couplings at low and high energies, then proceed with the calculation of corrections. Section~\ref{sec:pheno} is devoted to a discussion of the phenomenological consequences of these results. We summarise our introduced formalism and conclude in Section \ref{sec:summary}.


\begin{figure}[t!]
\centering
\includegraphics[width=0.22\textwidth]{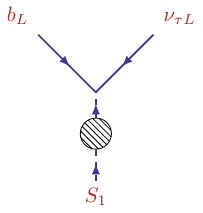}\hfill
\includegraphics[width=0.22\textwidth]{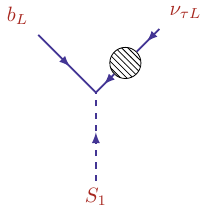}\hfill
\includegraphics[width=0.22\textwidth]{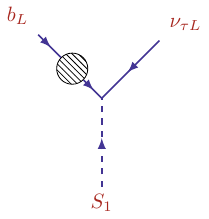}\\
\includegraphics[width=0.22\textwidth]{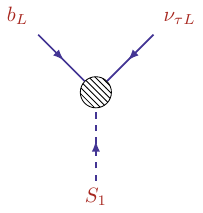}\hspace{1.2cm}
\includegraphics[width=0.22\textwidth]{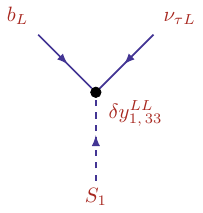}\hfill
  \caption{Radiative corrections to the $S_1$-$b_L$-$\nu_{\tau L}$ vertex ($y_{1\, 33}^{LL}$).
  The final diagram
  depicts the contribution from the coupling counterterm, $\delta y_{1\, 33}^{LL}$.}
  \label{fig:vertex}
  ~\\[-4mm]
  \hrule
\end{figure}

\section{Calculation}
\label{sec:setup}

The relevant Yukawa couplings of the LQ Lagrangian are given by
\begin{eqnarray}
  \mathcal{L}_{\rm LQ}
  &=&{\sum_{j,k=1}^3\sum_{a,b=1}^2}\Big[ 
      y_{1\, jk}^{LL} \,\overline{Q^c_L}{}^{j,a}\, \epsilon^{ab} L_L^{k,b} \,S_1 \;+\;
      y_{1\, jk}^{RR} \,\overline{u_R^c}{}^{j}  e_R^k \,S_1 \nn
  && \phantom{\sum_{j,k=1}^3\sum_{a,b=1}^2} -\; y_{2\, jk}^{RL} \,\overline{u_R}^j\epsilon^{ab} L_L^{k,b}\, R_2^a
      \;+\;
     y_{2\, jk}^{LR} \,   \overline{Q_L}{}^{j,a} e_R^k \,R_2^{a} \nn
   &&   \phantom{\sum_{j,k=1}^3\sum_{a,b=1}^2} 
     +\;   y_{3\, jk}^{LL} \,\overline{Q_L^c}^{j,a} \,\epsilon^{ab} L_L^{k,d}
      \,\lt( \sigma^l S_3^l \rt){}^{\! bd}   \; + \; \mbox{h.c.}\Big], \label{lag}
\end{eqnarray}
where we remark that our $ y_{2\, jk}^{LR}$ coincides with $ y_{2\, kj}^{LR*}$ of Ref.~\cite{Dorsner:2016wpm}. Here $Q^c = (u_L^c, d_L^c)$, the $\epsilon$ is the Levi-Civita tensor with $\epsilon^{12}=1$, and we do not consider a model with right-handed neutrinos. As the LQ masses exceed $1\tev$, we can neglect all fermion masses in calculations of corrections at $\mu_\text{high}$ and work in an SU(2) invariant basis. For completeness, we note that the quark and lepton doublets are defined in a basis in which the down-type fermions coincide  with the mass eigenstates, e.g.
\begin{eqnarray}
\label{eq:Q3def}
  Q^3 &=& \begin{pmatrix} V_{tb}^* t_L+V_{cb}^* c_L+V_{ub}^* u_L
    \\ b_L\end{pmatrix}, 
    \label{eq:dali}
\end{eqnarray}
with $V_{ij}$ referring to entries of the Cabibbo-Kobayashi-Maskawa (CKM) matrix.
{\eq{lag} is for one copy of $S_1$, $R_2$, and $S_3$ each. As mentioned in the introduction,
we will later consider several copies of $S_3$ carrying lepton flavor number, corresponding to the replacement $S_3\to S_{3k}$ in \eq{lag}.} An extension of the results in this section to other scalar LQ models may be found in Appendix~\ref{app:C}. 

To begin, we aim to calculate radiative corrections to LQ-fermion couplings responsible for the transitions $b\to s \ell^+\ell^-$ and $b\to c \tau \nu$. \fig{fig:vertex} shows these corrections
schematically for the case of $S_1$ coupling to left-handed $b$
quark and $\tau$ neutrino, $y_{1\, 33}^{LL}$. Here the left diagram illustrates self-energy
diagrams of the LQ involving quark-lepton loops. In processes that correspond to on-shell production of the LQ, this self-energy is
probed at $p^2=M_{S_1}^2$; in the low-energy 
observables
one is effectively probing the self-energy at $p^2=0$. The fermion self-energies (e.g. second-to-left diagram) have no $p^2$ dependence. The proper
vertex diagrams (third from the left) will also depend explicitly on $p^2$. We will now address how this energy-dependence may be utilized to define corrections to the LQ-fermion vertices.

\begin{figure}[t!]
\centering
\includegraphics[width=0.32\textwidth]{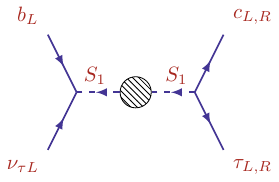}\hfill
\includegraphics[width=0.32\textwidth]{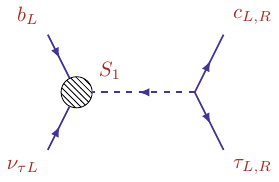}\hfill
\includegraphics[width=0.32\textwidth]{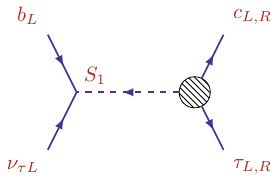}\hfill\\
\includegraphics[width=0.32\textwidth]{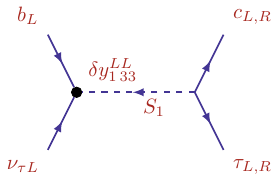}\hspace{0.6cm}
\includegraphics[width=0.4\textwidth]{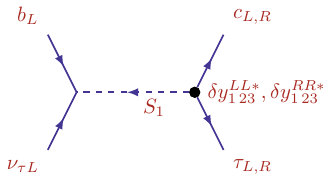}
  \caption{Radiative corrections to $b\to c\tau \nu$. 
    The fermion self-energy diagrams are not shown.
    $y_{1\, 33}^{LL,\rm low}$, $y_{1\, 23}^{LL,\rm low}$, and $y_{1\,
      23}^{RR,\rm low}$ are defined by choosing
    the coupling counterterm in the last two diagram to cancel the loop corrections.
  \label{fig:low} }
  ~\\[-2mm]
  \hrule
\end{figure}

First consider the loop corrected vertex for an on-shell
LQ (i.e.\ with $p^2=M_{S_1}^2$) and define an effective
\emph{high-energy coupling} by cancelling the correction by
a finite counterterm (rightmost diagram in \fig{fig:vertex}). This coupling
is the one probed in collider searches and used in the corresponding
event simulations. It can further be directly combined with QCD (or
other gauge) corrections, which are calculated in the usual way
by adopting the $\ov{\rm MS}$ scheme for the gauge coupling,
matching to effective field theories of four-fermion interactions
and utilising RGE techniques,
as detailed earlier. In an analogous way we define the \emph{low-energy coupling}
such that the radiative corrections to processes in \fig{fig:bdec}
vanish -- see \fig{fig:low} for the $S_1$ contribution to $b\to
c\tau\nu$. For the illustrated case of $y_{1\, 33}^{LL}$, the conversion factor $\kappa_{1\, 33}^{LL}\equiv
y_{1\, 33}^{LL,\rm high}/y_{1\, 33}^{LL,\rm low}$ will depend on the $S_1$ self-energies
evaluated at $p^2=M_{S_1}^2$ and  $p^2=0$, as well as the difference
of the proper vertex functions at $p^2=M_{S_1}^2$ and  $p^2=0$.
The fermion self-energies do not contribute to  $\kappa_{1\,33}^{LL}$ as they are not $p^2$ dependent.

Likewise $y_{1\, 33}^{LL,\rm low}$ is found from the diagrams in
\fig{fig:low}. Note that one can associate all corrections with
either of the two vertices (with the LQ self-energy in the first diagram 
shared between them) and therefore their cancellation by the finite
piece of the counterterms is well-defined. Since there are no 
box diagrams with two leptoquarks contributing to $b\to c\tau \nu$,
the renormalized one-loop correction  is zero when expressed
in terms of $y_{1\, 33}^{LL,\rm low}$. For this feature it is
essential that there are no box diagrams with two leptoquarks
contributing to $b\to c\tau \nu$. The situation is different for
$b\to s \ell^+\ell^-$ for which there can be box diagrams -- for instance
one can exchange two $S_1$ LQs if one permits couplings to $s$ quarks.
Thus the one-loop result for $b\to s \ell^+\ell^-$ expressed in terms of
$y_{1\, 33}^{LL,\rm low}$ should explicitly include these diagrams.
In the standard scenarios in which the dominant effect arises from
tree-level LQ exchange the smallness of the box loop makes these
corrections numerically irrelevant for the scope of this work.


\subsection{General pattern}
\label{sec:patt}

We now proceed with the discussion of the radiative corrections to the decay of a LQ produced on-shell at a collider and decaying into quark and lepton. We illustrate the calculation for
$y_{1\, {jk}}^{LL} \,\ov{Q^c_L}{}^{\, {j},a} \epsilon^{ab} L_L^{{k},b} \,S_1 $,
which drives the decays $S_1\to {\bar d_L^j \bar \nu_{L}^k}$ and
$S_1\to {\bar u{}_L^j \bar\ell_L^k}$ (and CKM suppressed channels). 
For on-shell fermions the tree-level $\ov{Q^c}{}^{j,a}$-$ L^{k,b}$-$S_1$ vertex amplitude reads 
\begin{equation}
A^{\rm tree} \; = \;\textit{\textit{}}
y_{1\, jk}^{LL}  \epsilon^{ab} \, \overline{u}_{j, b}{P_L}v_{k,a}, \label{atree}
\end{equation}
where $v_{k, a}$ ($u_{j, b}$) denotes the spinor of $L_k^b$ ($Q_j^{C,a}$) and there is no sum over $j,k$.
To demonstrate, we {first} consider $(j,k)=(3,3)$ {(i.e\ the left vertex in \fig{fig:low})}, the ratio of one-loop to tree amplitude for these decays is
\begin{eqnarray}
  \frac{A^{\rm one-loop}}{A^{\rm tree}}
  &=& 1 \;+\; \frac12 \delta Z_{S_1}  \;+\; \frac12 \delta Z_{Q}
      \;+\; \frac12 \delta Z_{L} \; +\; V_{L33}(M_{S_1}^2) \; +\;
      \frac{\delta y_{1\, 33}^{LL}}{y_{1\, 33}^{LL}}, \label{looptree}
\end{eqnarray}  
with the individual terms corresponding to the diagrams in
\fig{fig:vertex}. The proper vertex  correction $V_{L33}(p^2)$ depends on the
squared $S_1$ momentum $p^2$, which is equal to $M_{S_1}^2$
here. The $S_1$ wave function remormalization constant
(Lehmann-Symanzik-Zimmermann factor) $ \delta Z_{S_1}$ is related to the
bare self-energy $ \Sigma_{S_1}(p^2)$ as
\begin{eqnarray}
  \delta Z_{S_1}
  &=& \lt. {-} \frac{\partial}{\partial p^2}  \text{Re}\Sigma_{S_1}(p^2)
      \rt|_{p^2=M_{S_1}^2} . \label{lsz}
\end{eqnarray}  
The variables $\delta Z_Q$ and $\delta Z_L$ are the wave function constants of quark 
and lepton doublets, respectively.
The individual contributions in \eq{looptree} are ultraviolet (UV)
divergent and the UV pole of $\delta y_{1\, 33}^{LL}$ is chosen to
render $A^{\rm one-loop}/A_{\rm tree}$ finite. For the renormalization
scheme with $y_{1\, 33}^{LL}=y_{1\, 33}^{LL,\rm
  high}$ the counterterm $\delta y_{1\, 33}^{LL}$ receives a finite
piece such that the real part of the RHS of \eq{looptree} is equal to one,
\begin{eqnarray}
      \frac{\delta y_{1\, 33}^{LL,\rm high}}{y_{1\, 33}^{LL,\rm high}}
  &\equiv& \; \lt. \frac12 \,\frac{\partial}{\partial p^2}  \real \Sigma_{S_1}(p^2)
      \rt|_{p^2=M_{S_1}^2}  - \real V_{L33}(M_{S_1}^2) - \frac12 \delta Z_{Q}
      \;-\; \frac12 \delta Z_{L}.   
        \label{dyhigh}
\end{eqnarray}

The imaginary part of the vertex corrections does not contribute
to the decay rate $\propto |A^{\rm tree}+A^{\rm one-loop} |^2$ at
the considered one-loop order, if the {product of} couplings {in $ A^{\rm one-loop} $ has the same 
phase as $y_{1\, jk}^{LL}$ entering $A^{\rm tree}$.\footnote{If these phases are different one finds a non-zero CP asymmetry in the LQ decay, which is proportional to the imaginary part of the vertex function.}} 

Repeating this procedure for $y_{1\, 33}^{LL,\rm low}$ we first observe
that $ \Sigma_{S_1}(p^2)$ and $V_{L33}(p^2)$ enter the problem for $p^2=0$.
The $S_1$ self-energy in \fig{fig:low} is on an internal line and we
will absorb half of it into counterterms  for each of the adjacent
couplings. The blob in the
first diagram further includes the $S_1$ mass counterterm with Feynman rule
$-i \delta M_{S_1}^2$. For the usual ``pole mass'' definition $ \delta M_{S_1}^2$
is chosen as
\begin{eqnarray}
    \delta M_{S_1}^2 &=&  \real\Sigma_{S_1}( M_{S_1}^2), \no
\end{eqnarray}  
to ensure that the real part of the renormalized self-energy vanishes on-shell. 
Thus the ratio of one-loop to tree amplitude for
$b_L\to \bar \nu_{\tau L} c_L \tau_L$ is
\begin{eqnarray}
  \frac{A^{\rm one-loop}}{A^{\rm tree}}
  &=& 1 \;+\; \frac{\Sigma_{S_1}(0)-\real \Sigma_{S_1}(M_{S_1}^2) }{M_{S_1}^2} 
        \; +\; V_{L33}(0) \; +\; V_{L23}(0)^* \; +\;
      \frac{\delta y_{1\, 33}^{LL}}{y_{1\, 33}^{LL}} \; + \; 
      \frac{\delta y_{1\, 23}^{LL*}}{y_{1\, 23}^{LL*}} \nn
    &&  \; + \; 
      \mbox{fermion wave function constants}.
      \label{looptreelow}
\end{eqnarray}  
The factor of $M_{S_1}^2$ in the denominator of the second term stems
from the LQ propagator 
evaluated at $p^2=0$. 
The RHS of \eq{looptreelow} vanishes for the choice
\begin{eqnarray}
  \frac{\delta  y_{1\, 33}^{LL,\rm low}}{ y_{1\, 33}^{LL,\rm low}}
   &=& - \; \frac 12 \, \frac{\Sigma_{S_1}(0)-\real \Sigma_{S_1}(M_{S_1}^2)
       }{M_{S_1}^2} - V_{L33}(0) - \frac12 \delta Z_{Q}
      \;-\; \frac12 \delta Z_{L} , \label{dylow}
\end{eqnarray}
and the analogous definition of $\delta y_{1\, 23}^{LL,\rm low}$.  On
dimensional grounds one has $\Sigma_{S_1}(p^2) \propto p^2$ (for our
case of zero masses on internal lines), so that $\Sigma_{S_1}(0)=0$ and
the mass counterterm $\real \Sigma_{S_1}(M_{S_1}^2)$ dominates the
self-energy diagram in \fig{fig:low}. This is an important observation,
because $\Sigma_{S_1}(M_{S_1}^2)$ involves potentially sizable on-shell
loops, while $V_{L33}(0) $ involves small loop integrals evaluated at
zero external momenta. Low-energy physics probes heavy particles far
below their mass shell, therefore $\real \Sigma_{S_1}(M_{S_1}^2)$ must
enter $ y_{1\, 33}^{LL,\rm high}/ y_{1\, 33}^{LL,\rm low}$ in some way.
\footnote{One might ask whether it is allowed to set the top mass to zero in
$\Sigma_{S_1}(0)$, but keeping $m_t\neq 0$ leads to
$\Sigma (0)\propto m_t^2$ which is still negligible compared to
$\real \Sigma_{S_1}(M_{S_1}^2) \propto M_{S_1}^2$.} Combining
\eqsand{looptree}{dylow}, inserting \eq{lsz}, and generalising to an arbitrary 
fermion pair $(j,k)$ we arrive at
\begin{eqnarray}
  \kappa_{1\, {jk}}^{LL}
 &\equiv& \frac{y_{1\, {jk}}^{LL,\rm high}}{y_{1\,
  {jk}}^{LL,\rm low}} 
         \; = \;
        1 \;+ \; \frac{\delta y_{1\, {jk}}^{LL,\rm high}-\delta y_{1\, {jk}}^{LL,\rm
        low}}{y_{1\, {jk}}^{LL}} ,
        \nn
        &=&
       1 \,+ \frac12  \lt. \frac{\partial}{\partial p^2}  \text{Re}\Sigma_{S_1}(p^2)
        \rt|_{p^2=M_{S_1}^2} \!\!\!\! - 
        \frac 12 \, \frac{\real \Sigma_{S_1}(M_{S_1}^2) }{M_{S_1}^2}
           \, + V_{L{jk}}(0) \,-  {\real} V_{L{jk}}(M_{S_1}^2), \quad \label{kappa},
\end{eqnarray}  
which is a UV-finite quantity.
The fermion wave function renormalization constants cancel from $ \kappa_{1\, {jk}}^{LL}$, because they do not depend on the energy at which the leptoquark coupling is probed. This feature is important, because in models with several LQ copies, the diagrams with flavor-changing self-energies attached to the LQ-fermion coupling may come with couplings which are different from the tree-level coupling, permitting large corrections involving new parameters. The absence of these corrections therefore facilitates the phenomenological analyses.

\subsection{Self-energies}
\label{sec:self-e}

We find the fermion loop contribution to the bare $S_1$ self-energy as
\begin{eqnarray}
  \Sigma_{S_1} (p^2)
  &=& - \sum_{l,n=1}^3 \lt[ 2 |y_{1\ ln}^{LL}|^2 + |y_{1\  ln}^{RR}|^2
   \rt]
      \int \frac{d^D q}{(2\pi)^D i} \, 
      \frac{\mbox{tr}\,\lt[ q\sls (q\sls + p\sls) P_L\rt]}{q^2 (p+q)^2}
      \;\quad\mbox{with } P_L=\frac{1-\gamma_5}{2},\nn
  &=&   \sum_{l,n=1}^3 \lt[ 2 |y_{1\ ln}^{LL}|^2 + |y_{1\  ln}^{RR}|^2   \rt] p^2 (4\pi)^{-D/2}
      \, B_0(p^2; 0,0),
\end{eqnarray}
in dimensional regularization with $D=4-2\epsilon$ and  
\begin{eqnarray}
  B_0(p^2; m_1^2,m_2^2) &\equiv&
                                 \mu^{2\epsilon} \int \frac{d^Dq}{i \pi^{D/2}}
                                 \frac{1}{(q^2-m_1^2+i
                                 0^+)[(q+p)^2-m_2^2+i 0^+]}      ,
\end{eqnarray}
where $\mu$ is the renormalization scale.
One {then finds} that
\begin{eqnarray}
  B_0(p^2; 0,0) &=&
         \frac{1}{\epsilon} +2 -\gamma_E + i \pi -\ln \frac{p^2}{\mu^2}
          \;+\; {\cal O} (\epsilon),
\end{eqnarray}  
and, therefore
\begin{eqnarray}
  \Sigma_{S_1} (p^2)
  &=&\frac{  \sum_{l,n=1}^3\lt[ 2 |y_{1\ ln}^{LL}|^2 + |y_{1\  ln}^{RR}|^2 \rt]}{16\pi^2}
      \, p^2\, \lt[  \frac{1}{\epsilon} +2 -\gamma_E + \ln (4\pi) + i \pi -\ln
      \frac{p^2}{\mu^2} 
      \rt],
      \label{eq:sigma_S1}
\end{eqnarray}

\noindent where $\gamma_E$ is the Euler-Mascheroni constant. 
The self-energy contribution to $ \kappa_{1\, {jk}}^{LL}$ is therefore given by 
\begin{eqnarray}
  \kappa_{1\, {jk}}^{LL,\rm self}
  &=&  \frac12  \lt. \frac{\partial}{\partial p^2} \text{Re} \Sigma_{S_1}(p^2)
        \rt|_{p^2=M_{S_1}^2} \! - \,
        \frac 12 \, \frac{\real \Sigma_{S_1}(M_{S_1}^2) }{M_{S_1}^2},
\nn
  &=&
     - \frac{ \sum_{l,n=1}^3  \lt[ 2 |y_{1\ ln}^{LL}|^2 + |y_{1\  ln}^{RR}|^2 \rt]}{32\pi^2},\label{kappaself}
\end{eqnarray}
contributing universally to all $S_1$ couplings.

\begin{figure}[t!]
    \centering  \includegraphics{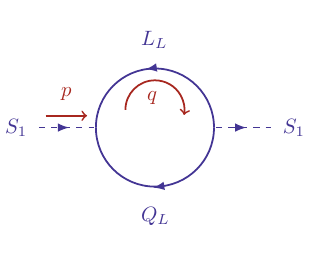}\hfill
\includegraphics{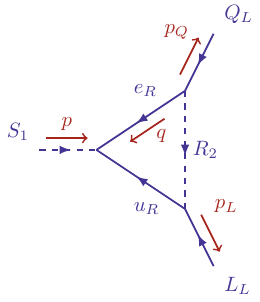}\hfill
\includegraphics{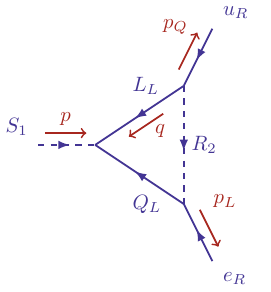}
    \caption{Diagrams for the loop correction to the $S_1$ LQ propagator (left) and  $S_1$ vertex corrections to the couplings $y^{LL}_1$ (middle) and $y^{RR}_1$ (right). }
    \label{fig:vertexCorrection}
      ~\\[-2mm]
  \hrule
\end{figure}

\subsection{Vertex corrections}
\label{sec:corrections}

In models which include additional LQs along with $S_1$, vertex corrections can involve an $R_2$ exchange. There are no such corrections with $S_3$ exchanged between the
external fermion lines. 

The one-loop vertex diagram  reads 
\begin{eqnarray}
  D_{Vjk} (p^2)&=&  \sum_{l,n=1}^3
   y_{2\, jl}^{LR*}     y_{1\, nl}^{RR}    y_{2\, nk}^{RL} \, \epsilon^{ab}\,
          (4\pi)^{-D/2} \label{dv} \\
          && \times \mu^{2\epsilon}   
     \int \frac{d^Dq}{i \pi^{D/2}}
   \frac{  \bar{u}_{{j}}[q\sls (q\sls +p \sls)] {P_L}v_{{k}} }{(q^2+i
                                 0^+)[(q+p)^2+i 0^+]
  [(q{+}p_Q)^2-M_{R_{{2}}}^2+i 0^+]},\nonumber 
\end{eqnarray}
and we note that
\begin{align}
    D_{Vjk}(p^2)
    {= V_{Ljk} (p^2) A_{\rm tree}},\label{dva}
\end{align}
with $A_{\rm tree}$ from \eq{atree}.
We use $q\sls (q\sls +p \sls) = q^2 +q\sls p\sls $ and define the
vector three-point function as 
\begin{eqnarray}
    C^\rho (p_2, p_3; m_1^2,m_2^2,m_3^2) 
  & \equiv& 
  \mu^{2\epsilon}
           \int \frac{d^D q}{i \pi^{D/2}}
            \frac{q^\rho}{[q^2-m_1^2][(q+p_2)^2-m_2^2][(q+p_3)^2-m_3^2]},
           \nn
  & \equiv & \phantom{+}
             p_2^\rho C_1(p_2^2, p_3^3, (p_2-p_3)^2; m_1^2, m_2^2, m_3^2) 
             \nn
       &&      +\;  p_3^\rho C_2 (p_2^2, p_3^3, (p_2-p_3)^2; m_1^2, m_2^2,
             m_3^2).
             \label{defcrho}
\end{eqnarray}
Using $p_2=p$,   $p_3=p_Q$, $p_2-p_3=p_L$, and $m_3=  M_{R_2}$
we find 
\begin{eqnarray}
  D_{Vjk} (p^2)&=&  \sum_{l,n=1}^3
   y_{2\, jl}^{LR*}     y_{1\, nl}^{RR}    y_{2\, nk}^{RL}
          \epsilon^{ab}\, 
         \,  (4\pi)^{-D/2} \label{dv2}. \\
          &&
          \times \lt[ B_0(0;0,M_{R_2}^2) + p^2 C_1 (p^2,0,0; 0,0,M_{R_2}^2)  \rt]
         [\overline{u}_{{j}}{P_L}v_{{k}}].
          \nonumber
\end{eqnarray}
$C_2$ does not contribute, because in \eq{dv} it comes with  $ \overline{u}_{a,j}
\slashed{p}_Q=0$ from the Dirac equation, where we neglect the mass of $Q$. For $p^2> 0$ one finds
\begin{eqnarray}
  &&C_1  (p^2,0,0; 0,0,M_{R_2}^2) \; \stackrel{\epsilon=0}{=}\; 
   \int_0^1 dy \int_0^{1-y} dx
     \frac{y}{-p^2 xy +m_{R_2}^2 (1-x-y) -i 0^+},  \nn
  &&= \; \frac{M_{R_2}^2}{p^4} \lt[
      \mbox{Li}_2\Big( - \frac{p^2}{M_{R_2}^2} \Big) +
      \lt( \ln  \frac{p^2}{M_{R_2}^2} - i \pi \rt)
      \lt( \ln \Big( 1+  \frac{p^2}{M_{R_2}^2} \Big) -
      \frac{p^2}{M_{R_2}^2} \rt) + \frac{p^2}{M_{R_2}^2} \rt].
      \qquad \label{c1res}
\end{eqnarray}
Since $p^2 C_1  (p^2,0,0; 0,0,M_{R_2}^2)  \to 0 $ for $p^2\to 0$,
only $ B_0(0;0,M_{R_2}^2)$ contributes to $D_{V\,jk}(0)$, which, however,
drops out of $ V_{Lln}(0) \,- \, V_{Lln}(M_{S_1}^2) $.
The other vertex corrections involve the same loop integrals and can be
found with trivial changes of the couplings.

\subsection{Final results}
\label{sec:Corrections}

Combining Eqs.~(\ref{kappa}), (\ref{kappaself}),
(\ref{dv}), (\ref{dva}), (\ref{dv2}) and (\ref{c1res}) 
we get our final result for the $S_1$ coupling to left-handed fermions:
\begin{eqnarray}
  \kappa_{1\, jk}^{LL}
  &=&   1 - \frac{ \sum_{l,n=1}^3  \lt[ 2 |y_{1\ ln}^{LL}|^2 +
      |y_{1\ ln}^{RR}|^2  \rt]}{32\pi^2} \;
   {-}
      \frac{\sum_{l,n=1}^3
   y_{2\, jl}^{LR*}     y_{1\, nl}^{RR}    y_{2\, nk}^{RL}}{y_{1\,
         jk}^{LL}\, 16\pi^2} f_\kappa \Big(  \frac{M_{S_1}^2}{M_{R_2}^2}
      \Big),\label{kllres}
\end{eqnarray}
with
\begin{eqnarray}
f_\kappa (x)  &=&
                   \frac1x \, \big[
      \mbox{Li}_2( - x ) +
      \ln  x \,  
      \lt[ \ln ( 1+ x ) - x \rt] + x      \big]  . \label{deff}
\end{eqnarray}
The function $f_\kappa (x)$ vanishes for $x=0$ and is positive for $0< x< 1.86$ with a maximum at $x=0.49$ and $f_\kappa(0.49)=0.23$. $f_\kappa (x)$ decreases monotonically for $x>0.49$ with $f_\kappa(3)=-0.24 $ and $f_\kappa(10)=-1.17$. 
  
By changing the couplings one immediately finds the other
scaling factors: 
\begin{eqnarray}
  \kappa_{1\, jk}^{RR}  &=& 1 - \frac{ \sum_{l,n=1}^3  \lt[ 2 |y_{1\ ln}^{LL}|^2 +
      |y_{1\ ln}^{RR}|^2  \rt]}{32\pi^2} \;
    {-2}  \; 
      \frac{\sum_{l,n=1}^3
   y_{2\, jl}^{RL}     y_{1\, nl}^{LL*}    y_{2\, nk}^{LR*}}{y_{1\,
         jk}^{RR*}\, 16\pi^2} f_\kappa \Big(  \frac{M_{S_1}^2}{M_{R_2}^2}
      \Big), \nn      
  \kappa_{2\, jk}^{LR}  &=&1 - \frac{ \sum_{l,n=1}^3  \lt[ |y_{2\ ln}^{LR}|^2 +
      |y_{2\ ln}^{RL}|^2  \rt]}{32\pi^2} \;
      -  \; 
      \frac{\sum_{l,n=1}^3
   y_{1\, jl}^{LL}     y_{2\, nl}^{RL*}    y_{1\, nk}^{RR*}}{y_{2\,
         jk}^{LR*}\, 16\pi^2} f_\kappa \Big(  \frac{M_{R_2}^2}{M_{S_1}^2}
      \Big),\nn
  \kappa_{2\, jk}^{RL}  &=& 1 - \frac{ \sum_{l,n=1}^3  \lt[ |y_{2\ ln}^{LR}|^2 +
      |y_{2\ ln}^{RL}|^2  \rt]}{32\pi^2} \;
      -  \; 
      \frac{\sum_{l,n=1}^3
   y_{1\, jl}^{RR*}     y_{2\, nl}^{LR}    y_{1\, nk}^{LL}}{y_{2\,
         jk}^{RL}\, 16\pi^2} f_\kappa \Big(  \frac{M_{R_2}^2}{M_{S_1}^2}
      \Big),\nn
  \kappa_{3\, jk}^{LL}  &=& 1 - \frac{ \sum_{l,n=1}^3   |y_{3\ ln}^{LL}|^2   }{16\pi^2}. \; \label{eq:correction}
\end{eqnarray}      

\subsubsection{\texorpdfstring{Conversion to $\mathbf{\overline{\rm MS}}$ scheme}{Conversion to MSbar scheme}}
\label{sec:MSbar}

\begin{figure}[t!]
  \centering \includegraphics{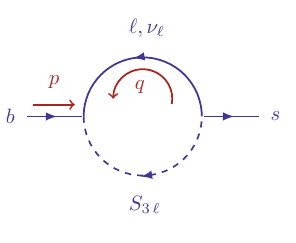}
  \caption{Diagram for the flavor-changing wave function renormalization factors induced by leptoquarks.\label{fig:btos-self} }
    ~\\[-2mm]
  \hrule
\end{figure}

LQs in the phenomenologically interesting mass range considered in this paper are best motivated as low-energy remnants of a theory with quark-lepton unification, which is typically realized at scales between ${\cal O}(10^3)\tev$ and the scale of 
grand unification (GUT scale). The construction of any viable model usually invokes the 
$\mathbf{\overline{\rm MS}}$ scheme for the lagrangian parameters, which are then evolved to 
lower scales with the RGE (see e.g. Ref.~\cite{Banik:2023ogi} for the 2-loop RGE in scalar LQ models). Also the IR fixed-points of the RGE for the LQ couplings found in 
Ref.~\cite{Fedele:2023rxb} correspond to the $\mathbf{\overline{\rm MS}}$ scheme. In this section we present the conversion of the LQ couplings between $\mathbf{\overline{\rm MS}}$ and our ``low" scheme. We exemplify this for the couplings of $S_{3\,\ell}$ for two reasons: 
First, in  Ref.~\cite{Fedele:2023rxb} non-trivial RG IR fixed-points were only found for 
this case and second, the popular $S_{3\,\ell}$ interpretation of $b\to s\ell^+\ell^-$ data 
implies that the considered LQ couples to both $b$ and $s$ and therefore permits flavor-changing $b$-to-$s$ self-energy diagrams. The proper treatment of the latter diagrams in the scheme conversion deserves some explanation. As a simplification, $S_{3\,\ell}$ only couples to doublet fermion fields, leading to a more compact expression compared to the cases with other LQs. 

The desired scheme conversion amounts to calculating the loop-corrected LQ-quark-lepton 
vertex and then determine the counterterm $\delta y_{3\, jk}^{LL}$ in both schemes,
the difference fixes $y_{3\, jk}^{LL,{\rm low}}-y_{3\, jk}^{LL,\overline{{\rm MS}}}$. 
In any scheme one has $y_{3\, jk}^{LL,{\rm bare}}=y_{3\, jk}^{LL,{\rm scheme}}+\delta y_{3\, jk}^{LL,{\rm scheme}}$, so that 
\begin{align}
y_{3\, jk}^{LL,\ov{\rm MS}} - y_{3\, jk}^{LL,{\rm low}} 
&=\; 
\delta y_{3\, jk}^{LL,{\rm low}} -\delta y_{3\, jk}^{LL,\ov{\rm MS}}.
\end{align}
In the  ``low'' scheme the counterterm $\delta y_{3\, jk}^{LL,{\rm low}}$ 
is given in terms of the vertex and self-energy loop functions 
in \eq{dylow}. 

The counterterm $\delta y_{3\, jk}^{LL,\ov{\rm MS}}$ can be obtained from \eq{looptree} 
by retaining only the pieces proportional to $1/\epsilon -\gamma_E+\ln (4\pi) $ of this equation.
Thus $\delta y_{3\, jk}^{LL,{\rm low}} -\delta y_{3\, jk}^{LL,\ov{\rm MS}}$ 
is simply found from the finite parts of the various loop contributions to 
$\delta y_{3\, jk}^{LL,{\rm low}}$ in \eq{dylow}. To this end we must calculate the 
wave function renormalization constants $\delta Z_Q$ and $\delta Z_L$ in both schemes,
which we did not need so far because they cancel from $\kappa_{3\, jk}^{LL}$.

We start with the discussion of $\delta Z_Q$, which is a matrix in flavor space. 
For the renormalization of the $\bar {b^c} \ell S_{3\,\ell}$ vertex we 
need $\left[ \delta Z_Q\right]_{33}$ and $\left[ \delta Z_Q\right]_{32}$ calculated from the 
$b\to b$ and $b\to s$ self-energies $\Sigma_{bb}\slashed{p}P_L$ and $\Sigma_{sb}\slashed{p}P_L$, respectively. $\Sigma_{sb}$ is shown in 
\fig{fig:btos-self}. It is calculated as 
\begin{align}
\Sigma_{sb}  \slashed{p}P_L \;=\; 
& - 3\sum_\ell y_{3\, 3\ell}^{LL} y_{3\, 2\ell}^{LL*} \, (4\pi)^{-D/2} \; \slashed{p} P_L \;  
   B_1(p^2,0,M_{S_{3\,\ell}}^2), \label{b1full} 
\end{align}
where
\begin{align} 
\label{b1full0} 
p^\nu B_1 (p^2,0,M_{S_{3\,\ell}}^2) \;=\; 
&    \mu^{2\epsilon} \int \frac{d^Dq}{i \pi^{D/2}}
                                 \frac{q^\nu}{(q^2+i
                                 0^+)[(q+p)^2-M_{S_{3\,\ell}}^2+i 0^+]}, \\
B_1 (0,0,M_{S_{3\,\ell}}^2) \;=\; & -\frac{1}{2\epsilon} -\frac14 +\frac12 \gamma_E +
\frac12 \ln \frac{M_{S_{3\,\ell}}^2}{\mu^2} \, +{\cal O} (\epsilon).\nonumber
\end{align}
Generalising the result to an arbitrary pair of quark fields, i.e.
$(s,b)\mapsto (a,j)$, we find the {hermitian} quark-field 
renormalization matrix
\begin{align}
 \delta Z_{Q, aj}^{\overline{\rm MS}} &= \frac{-3}{32\pi^2} \left( \frac{1}{\epsilon} -  \gamma_E + \ln(4\pi) \right) \sum_{k=1}^3 y_{3\, jk}^{LL} y_{3\, ak}^{LL*}. \label{eq:zq}
\end{align}
For the lepton self-energy, we only consider the case where each leptoquark copy $S_{3\ell}$
only couples to one lepton generation, which we label $n$,
\begin{align}
 \delta Z_{L, nn}^{\overline{\rm MS}} &= \frac{-3}{32\pi^2} \left( \frac{1}{\epsilon} -  \gamma_E +  \ln(4\pi) \right) \sum_{k=1}^3 \left| y_{3\, kn}^{LL}\right|^2.  \label{eq:zl}
\end{align}
Using \eq{looptree} with $S_1\to S_{3\ell}$ and \eqsto{eq:zq}{eq:zl} we find
\begin{align}
\delta y_{3\, jn}^{LL} 
&=\; 
-y_{3\, jn}^{LL} \left[ 
 \frac12 \delta Z_{S_3}   \;  +\; \frac12 \delta Z_{L\,nn}
 \right] 
\;-\; \frac12 \sum_{k=1}^3 \delta Z_{Q\, aj} \,y_{3\, an}^{LL},
   \label{eq:yct}
\end{align}
and therefore,
\begin{align}
\delta y_{3\, jn}^{LL,\overline{\rm MS}} 
=\; & 
\frac{1}{64\pi^2}\, \left( \frac{1}{\epsilon} -  \gamma_E +  \ln(4\pi) \right)\times  \label{eq:yctmsbar} \\ 
&\left[ 
   {4} \, y_{3\, jn}^{LL} \, \sum_{k=1}^3 \left| y_{3\, kn}^{LL}\right|^2 
\;+\; 3  \sum_{\ell,a=1}^3  y_{3\, j\ell}^{LL} \, y_{3\, a\ell}^{LL*} \,y_{3\, an}^{LL} \right].
\nonumber
\end{align}

The ``low" scheme is defined such that at zero momentum transfer all loop corrections to the LQ vertex 
are canceled by the counterterm, rendering the tree-level result exact. To calculate $ \delta y_{3\, jn}^{LL,\rm low} $ using \eq{dylow} we therefore also need the finite pieces of 
 $ \Sigma_{aj} $, distinguishing the cases $a=j$ and $a\neq j$. The flavor-conserving self-energy 
 $ \Sigma_{aa} $ enters    $ \delta y_{3\, jn}^{LL, \rm low} $ in the same way as in $\overline{\rm MS}$, but with the full result 
 for $B_1(0,0,M_{S_{3\ell}}^2)$ given in \eq{b1full0}.
 For $a\neq j$ one must instead calculate the diagram in which the tree-level $S_{3\, \ell}$ vertex is  amended by $ \Sigma_{aj} $ on the external quark leg 
 (upper right diagram in \fig{fig:vertex} with $S_1$ replaced by $S_{3\, \ell}$ and $\nu_{\tau L}$ 
 replaced by $L_\ell$). This calculation must be done keeping $m_a$ and $m_j$ non-zero, 
 the result will contain a non-trivial dependence on the quark mass ratio $m_j/m_a$. This feature is 
 familiar from e.g. the renormalization of the CKM matrix in the Standard Model~\cite{Denner:1990yz,Balzereit:1998id,Gambino:1998ec,Barroso:2000is,Diener:2001qt,Kniehl:2006bs}. 
The diagram evaluates to 
\begin{align}
   \bar u_j P_L v_n \sum_{\stackrel{\scriptstyle a=1}{a\neq j}}^{3} \frac{m_j^2}{m_j^2-m_a^2} y_{3\, an}^{LL} \Sigma_{aj} (p^2=m_j^2).
\end{align}
In this expression we can replace $\Sigma_{aj} (p^2=m_j^2) $ by $\Sigma_{aj} (0)$, because we discard sub-leading terms in $m_j^2/M_{S_{3\ell}}^2$. For the diagram with the lepton self-energy we 
do not encounter flavor-changing self-energies.

The final result for the desired counterterm reads  
\begin{align}
  \delta y_{3\, jn}^{LL,\rm low} 
  &= \delta y_{3\, jn}^{LL,\overline{\rm MS}}  
  - {\frac{1}{64\pi^2}} \left[ {6} \sum_{\stackrel{\scriptstyle a=1}{a\neq j}}^{3} \frac{m_j^2}{m_j^2-m_a^2}
   \sum_{\ell=1}^3 \left(\ln\frac{M_{S_{3 \ell}}^2}{\mu^2} -\frac12\right) 
          y_{3\, j\ell}^{LL} \, y_{3\, a\ell}^{LL*} \,y_{3\, an}^{LL} \right. \nonumber\\  
   &\hspace{6mm}       \left. + {3 \,y_{3\, jn}^{LL}}
         \sum_{\ell=1}^3 \frac{}{}\left( \ln\frac{M_{S_{3 \ell}}^2}{\mu^2} -\frac12\right)
          {\left|y_{3\, j\ell}^{LL}\right|^2}  
  + \, y_{3\, jn}^{LL} \, \left({7}\ln\frac{M_{S_{3 n}}^2}{\mu^2} -\frac{{19}}2\right)\sum_{a=1}^3 
  \left| y_{3\, an}^{LL}\right|^2   \right].
\end{align}
{We can define a scheme transformation factor as} 
\begin{align}
 \bar\kappa_{3\, jn}^{LL} \equiv& \, 
    \frac{y_{3\, jn}^{LL,\rm low}}{y_{3\, jn}^{LL,\overline{\rm MS}}} \,=\, 
    1 + \frac{\delta y_{3\, jn}^{LL,\rm low}- \delta y_{3\, jn}^{LL,\overline{\rm MS}}}{y_{3\, jn}^{LL}},
    \nonumber\\
    =& 1\, - {\frac{1}{64\pi^2}} \left[ {6} \sum_{\stackrel{\scriptstyle a=1}{a\neq j}}^{3} \frac{m_j^2}{m_j^2-m_a^2}
   \sum_{\ell=1}^3 \left(\ln\frac{M_{S_{3 \ell}}^2}{\mu^2} -\frac12\right) 
          \frac{y_{3\, j\ell}^{LL} \, y_{3\, a\ell}^{LL*} 
          \,y_{3\, an}^{LL}}{y_{3\, jn}^{LL}} \right. \nonumber\\  
   &\hspace{6mm}       \left. + {3}
         \sum_{\ell=1}^3 \frac{}{}\left( \ln\frac{M_{S_{3 \ell}}^2}{\mu^2} -\frac12\right)
          {\left|y_{3\, j\ell}^{LL}\right|^2} \,  
  +   \left({7}\ln\frac{M_{S_{3 n}}^2}{\mu^2} -\frac{{19}}2\right)\sum_{a=1}^3 
  \left| y_{3\, an}^{LL}\right|^2   \right] \label{eq:kb}.
\end{align}
A comment is in order here: The flavor-changing self-energies renormalising the 
coupling $y_{3\, jn}^{LL,\rm low}$ are calculated in the phase with broken electroweak 
SU(2) symmetry leading to the term $m_j^2/(m_j^2-m_a^2)$ in \eq{eq:kb}. This is unavoidable, 
because in the unbroken phase, with zero quark masses, one can arbitrarily rotate quark fields 
in flavor space and flavor quantum numbers are only well-defined in the broken phase. Therefore 
$\bar\kappa_{3\, jn}^{LL}$ should carry an index distinguishing down-type and up-type quarks.
In practice, however, the hierarchy among quark masses allows us to set either $m_a$ or $m_j$ to zero, so that  $m_j^2/(m_j^2-m_a^2)$ becomes 1 or 0, and $\bar\kappa_{3\, jn}^{LL}$ is the same for down-type and up-type quarks in this limit. In case one wants to keep the full dependence on 
quark masses in \eq{eq:kb}, $m_a$ and $m_j$ must be evaluated at the same value of the renormalization scale $\mu$. (Keep in mind that the ratio $m_j(\mu)/m_a(\mu)$ is 
$\mu$-independent.)  

Thus in the construction of a UV model of quark-lepton unification one will use $y_{3\, jn}^{LL,\overline{\rm MS}}$, subsequently run these couplings down to the scale defined by 
the $S_{3n}$ mass with the RG equations of Ref.~\cite{Fedele:2023rxb}, and finally convert the couplings to the ``low" scheme by multiplying them with $\bar\kappa_{3\, jn}^{LL}$ for applications in flavor physics. For use in collider 
physics one multiplies $y_{3\, jn}^{LL,\overline{\rm MS}}$ instead with $\bar\kappa_{3\, jn}^{LL} \kappa_{3\, jn}^{LL}$.

\section{Phenomenological implications}
\label{sec:pheno}

In order to demonstrate the impact of our results, we will outline phenomenological implications of the one-loop corrections calculated above in the context of flavor anomalies. To this end we will give numerical examples for the $\kappa$'s for values of the LQ couplings satisfying 
the constraints from the anomalous low-energy data. From \fig{fig:low} one realizes that the flavor observables constrain products of couplings while the $\kappa$'s in \eqsand{kllres}{eq:correction} instead depend on the individual couplings. Therefore the 
deviation of the $\kappa$'s from 1 is largest if one satisfies the low-energy constraints by choosing one coupling large and the other one small. This case is most relevant for 
collider searches of LQ, because at first, with little statistics, one probes the parameter region with small LQ masses and large couplings.
The complete one-loop corrections  to LQ decay rates involving  $y_{j\, ln}^{XY}$, where $XY=LL/LR/RL/RR$, are encoded in the $\kappa$'s, so that they can be considered pseudo-observables. We may study the size of the corrections, \textit{i.e.}\ the deviation of the $\kappa$'s from 1, to assess how large the couplings $y_{j\, ln}^{XY}$ can be without spoiling perturbation theory.

\subsection{Leptoquark decays}
\label{ssec:DecaySearches}
Here we consider the impact of these corrections for either explicit on-shell detection of a single LQ, or how signs of multi-LQ models may be extracted indirectly from single LQ decays. 
From the computation of the self energy above, we can derive the total decay width of each LQ as a function of their masses and couplings
\begin{align}
\Gamma_{S_1}&= \frac{M_{S_1}}{16\pi} \sum_{l,n} \left[ 2 |y_{1\ ln}^{LL}|^2 + |y_{1\  ln}^{RR}|^2 \right],\\
\Gamma_{R_2}&=\frac{M_{R_2}}{16\pi} \sum_{l,n} \left[ |y_{2\ ln}^{RL}|^2 + |y_{2\  ln}^{LR}|^2 \right],\\
\Gamma_{S_3}&=\frac{M_{S_3}}{8\pi} \sum_{l,n} |y_{3\ ln}^{LL}|^2.
\end{align} 
where the indices $l,n$ run over the different decay channels. By definition, the decay of LQs is determined by the high-energy couplings $y_{j\, ln}^{XY,high}$. The implication of the $\kappa$ factors that we computed appear as a re-scaling of a LQ decay width, which in turn, modifies the predictions for the branching ratios. For example, a third-generation $S_1$ LQ will have modified branching ratios, for example
\begin{align}
\text{BR}(S_1 \to b\nu_\tau)
   &=\frac{1}{\Gamma_{S_1}}\frac{M_{S_1}}{16\pi} |y_{1\, 33}^{LL,\mathrm{high}}|^2, \\ \nonumber
   &= \frac{|y_{1\, 33}^{LL,\mathrm{low}}|^2\times|\kappa_{1\, 33}^{LL}|^2}{\sum_{l,n} \left[ 2 |y_{1\ ln}^{LL,\mathrm{low}}|^2|\kappa_{1\, ln}^{LL}|^2 + |y_{1\  ln}^{RR,\mathrm{low}}|^2|\kappa_{1\, ln}^{RR}|^2 \right]}, \\
\text{BR}(S_1 \to \tau t)
   &=\frac{1}{\Gamma_{S_1}}\frac{M_{S_1}}{16\pi} (|y_{1\, 33}^{LL,\mathrm{high}}|^2+|y_{1\, 33}^{RR,\mathrm{high}}|^2), \\ \nonumber
   &= \frac{(|y_{1\, 33}^{LL,\mathrm{low}}|^2|\kappa_{1\, 33}^{LL}|^2+|y_{1\, 33}^{RR,\mathrm{low}}|^2|\kappa_{1\, 33}^{RR}|^2)}{\sum_{l,n} \left[ 2 |y_{1\ ln}^{LL,\mathrm{low}}|^2|\kappa_{1\, ln}^{LL}|^2 + |y_{1\  ln}^{RR,\mathrm{low}}|^2|\kappa_{1\, ln}^{RR}|^2 \right]},
\end{align}
where we have neglected phase space effects, as $m_t^2/m_{S_1}^2$ can barely exceed 2\%. In order to calculate the impact of a model built for low scale observables on the direct production rates at colliders then one should incorporate these radiative corrections. 

To determine the hierarchy between the couplings, one can measure the ratio of branching ratios, for example
\begin{align}
    \frac{\text{BR} (S_1 \to \tau t)}{\text{BR} (S_1 \to \nu_\tau b)} &\approx \frac{|y_{1\, 33}^{LL,\mathrm{high}}|^2+|y_{1\, 33}^{RR,\mathrm{high}}|^2}{|y_{1\, 33}^{LL,\mathrm{high}}|^2}.
\end{align}
From this observable we note the following relation between low- and high-energy couplings
\begin{align}
\frac{|y_{1\, 33}^{LL,\mathrm{high}}|^2+|y_{1\, 33}^{RR,\mathrm{high}}|^2}{|y_{1\, 33}^{LL,\mathrm{high}}|^2} = \frac{|y_{1\, 33}^{LL,\mathrm{low}}|^2|\kappa_{1\, 33}^{LL}|^2+|y_{1\, 33}^{RR,\mathrm{low}}|^2|\kappa_{1\, 33}^{RR}|^2}{|y_{1\, 33}^{LL,\mathrm{low}}|^2|\kappa_{1\, 33}^{LL}|^2}.
\end{align}
In the absence of an additional LQ $R_2$, it follows from Eqs.~\eqref{kllres} and \eqref{eq:correction} that $\kappa_{1\, 33}^{RR} =\kappa_{1\, 33}^{LL}$, i.e. the radiative corrections to the right- and left-handed couplings are equal and
\begin{align}
\frac{\text{BR} (S_1 \to \tau t)}{\text{BR} (S_1 \to \nu_\tau b)} &\approx \frac{|y_{1\, 33}^{LL,\mathrm{low}}|^2+|y_{1\, 33}^{RR,\mathrm{low}}|^2}{|y_{1\, 33}^{LL,\mathrm{low}}|^2},
\end{align}
which suggests that the ratio of couplings is insensitive to radiative corrections. However, the presence of $R_2$ may break this relation, which implies that this ratio can act as an indirect probe of the presence of a second LQ with sizeable couplings.

\subsection{Flavour anomalies}
\label{ssec:FlavourAnomalies}

Several measured branching ratios and decay distributions have hinted at the presence of BSM physics influencing the branching ratios of $b\to s \ell \ell$ and $b\to c \ell \nu$. Notably in the neutral current process these are driven by the decay $b\to s \mu\mu$, where there is a deficit in events in the kinematic region $q^2\leq 8$ GeV \cite{LHCb:2014cxe,LHCb:2015wdu,LHCb:2021zwz} ($q^2$ is the invariant mass of the lepton pair), in contrast with the SM predictions of Refs.~\cite{Khodjamirian:2010vf, Khodjamirian:2012rm}. The angular observable $P_5'$, which parameterizes the angular distribution of final states in $B\to K^* \mu^+\mu^-$, also follows this trend~\cite{LHCb:2013ghj,LHCb:2015svh,LHCb:2020lmf,LHCb:2020gog}. Measurements of the lepton flavor universality~(LFU) ratios~\cite{Hiller:2003js}
\begin{align}
R_{K^{(*)}} = \frac{\text{BR}(B\to K^{(*)} \mu^+\mu^-)}{\text{BR}(B\to K^{(*)} e^+ e^-)} \label{eq:RKstar},
\end{align}
presently show a compatibility with the SM predictions $R_{K^{(*)}} \simeq 1$~\cite{LHCb:2022qnv, LHCb:2022vje}, indicating that the BSM physics invoked to explain the anomalous data in $b\to s \mu^+\mu^-$ should couple with similar strengths to electrons and muons. For the charged-current process, a long-standing anomaly is observed in the LFU ratios 
\begin{align}
    R_{D^{(*)}} = \frac{\text{BR}(B\to D^{(*)} \tau \nu)}{\text{BR}(B\to D^{(*)} \ell \nu)}, \quad \ell = e, \mu.\label{eq:defrd}
\end{align}
The HFLAV collaboration combines six separate measurements from BaBar~\cite{BaBar:2012obs}, Belle~\cite{Belle:2015qfa,Belle:2016ure,Belle:2019rba},  and LHCb~\cite{LHCb:2023zxo,LHCb:2023uiv} to give the following average values, as per Summer 2023~\cite{HFLAV:2022esi}
\begin{align}
    R_D^\text{HFLAV}= 0.357 \pm 0.029,\qquad \;\;\;\;  R_{D^*}^\text{HFLAV}= 0.284 \pm 0.012 ,\label{eq:HFLAVbctaunu}
\end{align}
with a correlation coefficient of $\rho=-0.37$. To be compared with the theoretical values
\begin{align}
    R_D =0.298\pm 0.004,\qquad \;\;\;\;R_{D^*} =0.254\pm 0.005,
\end{align}
resulting in a theory/experiment discrepancy of $3.2\sigma$

Improved sensitivity to the $D^*$ and $\tau$ polarizations can discriminate between different BSM explanations of these anomalies, and complementary information can be garnered from the ratio $R_{\Lambda_c}\equiv \text{BR}(\Lambda_b \to \Lambda_c \tau \nu)/\text{BR}(\Lambda_b \to \Lambda_c \ell \nu)$~\cite{Blanke:2018yud,Blanke:2019qrx}. For a BSM explanation for $R_{D^{(*)}}$ to be consistent with the latter, future measurement of $R_{\Lambda_c}$ must move upwards from the 2022 value $R_{\Lambda_c}^\text{LHCb}= 0.242\pm 0.026\pm 0.040\pm 0.059$~\cite{LHCb:2022piu} to $R_{\Lambda_c}=0.39\pm 0.05$~\cite{Fedele:2022iib}, to be consistent with Eq.~\eqref{eq:HFLAVbctaunu}.

These flavor anomalies may be addressed by a combination of the scalar LQs $S_1, R_2$ and $S_3$ (a $SU(2)$ singlet, doublet and triplet, respectively). The LQs $S_1$ and $R_2$ are capable of explaining the flavor anomalies with tree-level contributions to $b\to c\tau\nu$, and $S_3$ to $b\to s\mu\mu$. In order to be consistent with LFU, i.e. Eq.~\eqref{eq:RKstar}, one requires multiple copies of $S_3$ in order to consistently explain $b\to s\ell\ell$ data and to avoid constraints from lepton flavor violating decays, e.g. $\mu\to e \gamma$. Therefore this motivates extensions to the SM involving the combinations $(S_1, S_{3\,\ell})$ or $(R_2, S_{3\,\ell})$, each with at least two copies of $S_{3}$ of the triplet LQ (denoted $S_{3\,\ell}$) with $S_{3\,e}$ and $S_{3\,\mu}$ exclusively coupling to $e$ and $\mu$, respectively. We do not perform any phenomenological analysis of constraints on these models here, but simply {use the best-fit values of the Wilson coefficients from the literature to employ realistic scenarios for the LQ couplings and masses in our numerical examples.}  We draw the readers' attention to Ref.~\cite{Fedele:2023rxb} and references therein for further discussion of constraints.

\subsubsection{Scalar singlet}
\label{sec:singlet}
As discussed above, the singlet $S_1$ is capable of generating tree-level contributions to $b\to c\tau\nu$. {We} adopt a minimal coupling scenario for the $S_1$ scalar LQ {choosing} nonzero values only {for the} couplings $y_{1\, 33}^{LL}$ and $y_{1\, 23}^{RR}$. The contributions to $b\to c \ell \nu_\ell$ are given by the following effective interactions, expressed in the \texttt{flavio} basis \cite{Straub:2018kue} for the Weak Effective Theory (WET), where \begin{align}
\label{eq:WET}
    \mathcal{L} = \mathcal{L}_\text{SM} -\left[\frac{4G_F}{\sqrt{2}}V_{cb}\sum_{b\to c \tau\nu} C_i \mathcal{O}_i\, \rm{\textcolor{blue}{+ h.c.}}\right],
\end{align}   
    with the sum indicating a sum over the operator basis for the process $b\to c \tau\nu$. The operators of interest are
\begin{align}
\label{eq:EFTbasis}
&\mathcal{O}_{S_L} = (\overline{c} P_L b)(\overline{\tau}P_L \nu_\tau),\quad \quad
\mathcal{O}_{T}= (\overline{c} \sigma^{\mu\nu} P_L b)(\overline{\tau}\sigma_{\mu\nu}P_L\nu_\tau),\\
&\mathcal{O}_{V_L} = (\overline{c}\gamma^\mu P_L b)(\overline{\tau}\gamma_\mu P_L\nu_\tau). \nonumber
\end{align}
Here, $G_F$ is the Fermi constant, and $V_{ij}$ is the SM CKM matrix elements. Note that here we have restricted ourselves to lepton-flavor conserving effective interactions. This minimal scenario will generate the following nonzero WCs, where we have assumed the LQ couplings to be real-valued
\begin{align}
\label{eq:S1WCExpressions}
C_{S_L} (M_{S_1}) &= -4 C_{T}(M_{S_1}) = -\frac{1}{4\sqrt{2}G_FV_{cb} M_{S_1}^2} y_{1\, 23}^{RR} y_{1\, 33}^{LL} ,\\
C_{V_L}(M_{S_1}) 
&= \frac{1}{4\sqrt{2}G_F M_{S_1}^2}(y_{1\, 33}^{LL})^2 .
\end{align}
We recall we have adopted the down-aligned mass basis for the quarks, see Eq.~\eq{eq:dali}. These LQ couplings can be chosen to be fixed in the low-scale renormalization scheme.

A recent fit to these WCs in a model-dependent study of the anomalies $R_{D^{(*)}}$ and other $b\to c\tau\nu$ observables can be found in Ref.~\cite{Iguro:2022yzr}. Their results yield the following best fits to single scalar/tensor WCs, evaluated at $\mu_b$,
\begin{align}
    C_{S_L}= -8.9 C_{T}=0.19
    \label{eq:Vector-scalarfit}
\end{align}
which, when evolved according to Appendix~\ref{app:A}, corresponds to $C_{S_L}=-4 C_T \approx 0.095$ at $M_{S_1}=2$ TeV. Although this fit was performed prior to the most recent updates to the $R_{D^*}$ measurements, it remains compatible with the central values of the present HFLAV fit~\cite{HFLAV:2022esi} to these anomalies within roughly $1.3\sigma$.  Matching onto Eq.~\eqref{eq:S1WCExpressions}, this corresponds to
\begin{align}
    { y_{1\, 23}^{RR}\, y_{1\, 33}^{LL}}\approx-1.06\;
\end{align}
for a LQ mass of $2$ TeV.

A fit to the anomalous $R_{D^*}$ measurements arising solely from the $C_{V_L}$ in the $S_1$ model is ruled-out primarily due to large contributions to $b\to s \nu\bar{\nu}$ (see Appendix~\ref{app:B}) which are ruled-out by present measurements. In the framework prescribed above, it is impossible to avoid nonzero contributions to the $C_{V_L}$ coefficient, although it is CKM suppressed. For this reason, we elect for a larger value for $y^{RR}_{1\, 23}$ to minimize the effect from the vector WC. Dominant constraints on this coupling come from high-$p_T$ probes of the effective coupling in $c\bar{c}\to \tau\bar{\tau}$ interactions, resulting in a constraint $|y^{RR}_{1\, 23}|<2.6$ for a $2$ TeV leptoquark mass~\cite{Angelescu:2018tyl, Bigaran:2022kkv}. Saturating this upper bound corresponds to a candidate point  $(y_{1\, 23}^{RR},y_{1\, 33}^{LL})=(2.6, -0.41)$. For this point, $C_{V_L}$ has negligible impact on the values for $R_{D^{(*)}}$, meaning that this point is still able to fit the anomalies while being safe from constraints. Matching onto Eq.~\eqref{eq:S1WCExpressions}, this point is found to yield a value of 
\begin{align}
     \kappa_{1\, jk}^{LL}= \kappa_{1\, jk}^{RR} = 1- \frac{2|y^{LL}_{1\, 33}|^2 + |y^{RR}_{1\, 23}|^2}{32\pi^2} \approx 0.98,
\end{align}
and in the scalar singlet LQ model to explain the anomalies in $b\to c\tau\nu$ this correction corresponds to a $2\%$ effect. Therefore, the high energy Yukawa coupling relevant for on-shell production of the $S_1$ LQ is 98\% of the value of that at low-energy.

\subsubsection{Scalar doublet}
\label{scalardoublet}

Following similarly from Eqs.~\eqref{eq:EFTbasis} and ~\eqref{eq:WET}, the doublet $R_2$ generates the following nonzero WCs for contributing to the $b\to c\tau\nu$ process, this time not assuming real-valued LQ couplings~\footnote{In a model with two distinct $R_2$ doublets, one may also generate a contribution to a right-handed vector coefficient, $C_{V_R}$ (see e.g. Ref.~\cite{Endo:2021lhi}).}
\begin{align}
 \label{eq:R2WCExpressions}
C_{S_L} (M_{R_2}) &= 4 C_{T}(M_{R_2}) = \frac{1}{4\sqrt{2} G_F V_{cb} M_{R_2}^2} y^{LR\; *}_{2\; 33} y^{RL}_{2\;23}.
\end{align}
Following again from the fit in Ref.~\cite{Iguro:2022yzr},  the following best fit to single scalar/tensor WCs, evaluated at $\mu_b$,
\begin{align}
    C_{S_L}= 8.4 C_{T}\approx -0.07 \pm 0.58\,i ,
    \label{eq:Vector-scalarfit2}
\end{align}
which, when evolved according to Appendix~\ref{app:A}, corresponds to $C_{S_L}\approx 8.4 C_T =-0.039 \pm 0.319\,i$ at $M_{R_2}=2$ TeV. Matching onto Eq.~\eqref{eq:R2WCExpressions}, this corresponds to
\begin{align}
   { y^{LR\; *}_{2\; 33} y^{RL}_{2\;23}}\approx-0.41\pm 3.43\,i\;,
\end{align}
again for a LQ mass of $2$ TeV. Dominant constraints on these couplings arise from a loop-order corrections to $Z \to \tau\overline{\tau}$ couplings and the high-$p_T$ dilepton and monolepton tails of proton collisions. 

The light-quark couplings to taus are limited by constraints on the flavor-con\-ser\-ving, non-universal, contact interactions in $pp\to \tau^+\tau^-$ and $pp\to \tau \nu$ (+ soft jets)\cite{Faroughy:2016osc,Greljo:2017vvb,Angelescu:2018tyl}. Utilising the program \texttt{HighPT} \cite{Allwicher:2022mcg,Allwicher:2022gkm} to extract these constraints and mapping the result onto the above model, we find that requiring agreement at 3(2)$\sigma$ constrains $|C_{S_L}|\lesssim 0.68(0.57)$ for a $2$ TeV LQ mass. This is consistent with Ref.~\cite{Becirevic:2024pni} where they examine the $R_2$ model and conclude that these constraints are incompatible with an $R(D^{(*)})$ if the high-$p_T$ constraints are taken at $2\sigma$, although they note that these constraints should be treated with caution due to uncertainties from $\tau$ reconstruction~\cite{Jaffredo:2021ymt}. We conservatively take the $3\sigma$ constraint and retain the above point as valid, as assessing the definite validity is beyond the scope of this work.

For $Z \to \tau\overline{\tau}$, the coupling of the LQ to the tau and top quarks $\approx y^{LR}_{2\,33}$ generates a enhanced correction~\cite{Becirevic:2018uab}. A representative coupling assignment which satisfies this constraint (see Appendix~\ref{app:B}) is $( y^{LR}_{2\; 33},y^{RL}_{2\;23})\approx ( 0.5,-0.82\pm 6.86\,i)$, corresponds to
 \begin{align}
     \kappa_{2\, jk}^{LR}= \kappa_{2\, jk}^{RL}=  1- \frac{|y^{LR}_{2\, 33}|^2 + |y^{RL}_{2\, 23}|^2}{32\pi^2} \approx0.85. \label{eq:ex85}
 \end{align}
So, for the scalar doublet LQ model to explain the anomalies in $b\to c\tau\nu$ this correction corresponds to a $15\%$ effect and the high energy Yukawa coupling relevant for on-shell production of the $R_2$ LQ is 85\% of the value of that at low-energy.
Of course the large deviation from 1 in \eq{eq:ex85} stems from our choice with $|y^{RL}_{2\;23}|\gg |y^{LR}_{2\; 33}|$. If  chose $|y^{RL}_{2\;23}|\approx |y^{LR}_{2\; 33}|\approx 1.86$ instead, we'd find $ \kappa_{2\, jk}^{LR}= \kappa_{2\, jk}^{RL}=0.98$ instead. Yet, as stated at the beginning of this section, collider searches first probe the large-couplings region and \eq{eq:ex85} shows that the corrections encodes in our $\kappa$'s matter here. 
We also verify that LQ couplings as large as seven do not lead to excessive corrections which would put the validity of perturbation theory into doubt.

\subsubsection{Scalar triplet}
\label{sec:triplet}

The scalar triplet is relevant for addressing anomalies in $b\to s \ell\ell$. The effective Lagrangian parameterizing these transitions is given by
\begin{align}
    \mathcal{L}= \mathcal{L}_\text{SM} - \frac{4 G_F}{\sqrt{2}} V_{tb} V_{ts}^* \left(C_9^\ell \mathcal{O}_9^\ell + C_{10}^\ell \mathcal{O}_{10}^\ell \right) +\text{h.c.},
\end{align}
and the operators are given by
\begin{align}
    \mathcal{O}^\ell_9= \frac{\alpha_\text{em}}{4\pi} \left( \overline{s} \gamma_\mu P_L b\right)\left(\overline{\ell} \gamma^\mu \ell \right),\;\;\; \;\;\mathcal{O}_{10}^\ell=\frac{\alpha_\text{em}}{4\pi} \left( \overline{s} \gamma_\mu P_L b\right)\left(\overline{\ell} \gamma^\mu \gamma_5 \ell \right).
\end{align}
Here, $\alpha_\text{em}$ is the fine structure constant.

Matching the scalar triplet LQ model onto this Lagrangian gives the following contribution
\begin{align}
C^\ell _9(M_\text{LQ}) =-C^\ell _{10}(M_\text{LQ}) =- \frac{\pi }{2\sqrt{2} G_F V_{tb} V_{ts}^* \alpha_\text{em}} \frac{y^{LL}_{3\, 3\ell}y^{LL\; *}_{3\, 2\ell}}{M_{S_3}^2}.
\end{align}
For the triplet LQ it is useful to define the quantity $C^\ell_L= C_9^\ell= - C_{10}^\ell$, and recent global fits show a preference for a universal assignments of these WCs such that $C^U_L=C^e_L=C^\mu_L=C^\tau_L\approx -0.4$ \cite{Ciuchini:2022wbq,Greljo:2022jac,Alguero:2023jeh}. \footnote{Note that there is a preference for $C_{10}^\mu=0$, i.e. the SM case, {in the data on} the branching ratio $B(B_s\to \mu\mu)$ if {the SM prediction is calculated with $V_{cb}$ determined from \emph{inclusive} $b\to c\ell\nu$ decays, since then the SM prediction is close to the central value of the measurement. \emph{Exclusive} $b\to c\ell\nu$ decays prefer a smaller 
$|V_{cb}|$, so that $C_{10}^\mu \neq 0$ improves 
the agreement with $B(B_s\to \mu\mu)$ 
(see e.g. \cite{Lang:2022mxu}).In any case a combined fit of all data prefers a scenario with 
$C_{9}^\ell=-C_{10}^\ell \neq 0$ over the SM case.}}

We begin by considering a single triplet, $S_3$, with lepton-universal couplings. Recalling that $C^U_{L}$ as a vector coefficient does not run in QCD, this
corresponds to 
\begin{align}
y^{LL}_{3\, 3\ell}y^{LL\; *}_{3\, 2\ell} \approx -0.0055 [-0.14],
\end{align}
for a LQ mass of $2$ TeV [10 TeV]. Note that for the corrections to the Yukawa couplings $\kappa^{LL}_{3, \,jk}$ are generated uniformly in flavor space $(j,k)$ because they are each sourced by the LQ self-energy. For $\kappa^{LL}_{3, \,jk}$ to correspond to a $30\%$ effect relies on a coupling assignment for a LQ mass of $2$ TeV of $(y^{LL}_{3\, 
3\ell},y^{LL}_{3\, 2\ell})\approx (6.1, -10^{-3})$ 
and for a LQ mass of $10$ TeV of $(y^{LL}_{3\, 3\ell},y^{LL}_{3\, 2\ell})\approx (6.1, -0.023)$. Therefore in each of these instances the high-energy, lepton-universal, Yukawa coupling relevant for on-shell production of the $S_3$ LQ is 70\% of the value of that at low-energy.

\subsubsection{A comment on models with multiple LQ}
\label{sec:multipleLQ}

The models considered in Section~\ref{ssec:FlavourAnomalies} all involve the addition of one LQ at a time and so their corrections are dominated by the LQ self-energies. However, the results from Section~\ref{sec:Corrections} illustrate that the largest and most interesting effects will occur when two LQs are present in a model: namely, in models with both $S_1$ and $R_2$ present. In this model, the Lagrangian in Eq.~\eqref{lag} reduces to 
\begin{eqnarray}
  \mathcal{L}_{\rm LQ}
  &=& \phantom{+\;} 
      y_{1\, jk}^{LL} \,\overline{Q^c_L}{}^{j,a}\, \epsilon^{ab} L_L^{k,b} \,S_1 \;+\;
      y_{1\, jk}^{RR} \,\overline{u_R^c}{}^j  e_R^k \,S_1 \nn
  && -\; y_{2\, jk}^{RL} \,\overline{u_R}^j\epsilon^{ab} L_L^{k,b}\, R_2^a
     \quad \;+\;
     y_{2\, jk}^{LR} \,   \overline{Q_L}{}^{j,a} e_R^k \,R_2^{a} \; + \; \mbox{h.c.}.
\end{eqnarray}
The LQs $S_1$ and $R_2$ are particularly phenomenologically interesting because they are the sole two mixed-chiral scalar LQs: they have couplings to both right- and left-handed quarks and leptons. These leptoquarks have been exploited in the literature, for example, to generate large contributions to chirality-flipping dipole operator corrections, responsible for radiative mass generation and large dipole moments for charged leptons (see e.g. Refs.~\cite{Bigaran:2020jil,Bigaran:2021kmn}). Nevertheless, these models do not necessarily require both LQs at once. However, future measurements of lepton CLFV and associated processes may necessitate such model building in the near future.

A particularly interesting feature of the results in Section~\ref{ssec:FlavourAnomalies} is an inverse dependence of the vertex correction on the low-scale LQ coupling being corrected. This means that even a small low-scale coupling can have a large correction,
simply because the tree-level coupling is suppressed or even absent in the loop vertex diagram, which involves three different couplings instead.
Take for example a LQ model with both $S_1$ and $R_{2}$ and consider the correction to the coupling $y_{1\, jk}^{RR*}$, taking LQ masses such that $f({M_{S_1}^2}/{M_{R_2}^2}=10)=-1.17$. Fixing each of the LQ couplings to third-generation quarks to $0.1$ aside from the LQ coupling we are correcting for the purpose of demonstration\footnote{Fixing the ratio of masses but allowing flexibility for the absolute masses may allow one to suppress phenomenological impact of this choice.},
\begin{align}
\kappa_{1\, jk}^{RR}  = 1 -3\,|y_{1\,jk}^{RR*}|^2\times 10^{-3}
    + \; \frac{4.4\times 10^{-5}}{y_{1\,jk}^{RR*}}.
\end{align}
Therefore, for a LQ coupling $y_{1\,jk}^{RR*}\approx 10^{-4}$ the correction is $\kappa_{1\, jk}^{RR} \approx 1.44$, corresponding to a correction of $44\%$ to this coupling.

\subsection{Unification and IR fixed-points}
\label{ssec:unification}

Leptoquarks particularly appear in theories that unify quarks and leptons, although the unification scale $M_\text{QLU}$ is often many orders of magnitude above the LQ mass scale. If we assume that the mass-gap between the electroweak scale and $M_\text{QLU}$ is populated only by LQs and SM particles, one may use renormalization-group evolution of LQ couplings to fermions up to this scale.  This analysis was performed in Ref.~\cite{Fedele:2023rxb}. The scale $M_\text{QLU}$ determines the masses of any remaining particles in the unification framework, and the effects of these particles decouple for a scalar LQ model as $M_\text{QLU} \to \infty$ \footnote{For a vector LQ model where $M_\text{LQ}\ll M_\text{QLU}$ this corresponds to a non-decoupling scenario unless a Higgs sector responsible for the LQ mass is also considered~\cite{Fedele:2023rxb}.}.

In Ref.~\cite{Fedele:2023rxb} it has been found that the Yukawa couplings of a model with three scalar triplet LQ carrying lepton flavor number features IR fixed-points withemergent lepton flavor universality in their RG evolution from some high scale down to $M_{LQ}$. This feature persists if  additional LQs (such as $S_1$) are present. Since beta functions are derived in mass independent renormalization schemes, typically $\overline{\mathrm{MS}}$, 
the couplings must be converted to the `low' or `high' scheme for use in phenomenological applications.

Taking $M_{S_{3}} {\equiv} M_{S_{3\,e}} = M_{S_{3\, \mu}}= M_{S_{3\, \tau}} {=} 14.2$ TeV  and the IR fixed-point couplings  in Table~\ref{tab:LQU}.\footnote{These points are from Table 6 of Ref.~\cite{Fedele:2023rxb}, where our notation corresponds to $y^{LL}_{3\, ij}=y^{e}_{3\, ij}$. } reproduces the $b\to s \ell^+\ell^-$ data. In each case the conversion between low-energy and high-energy couplings is given by
\begin{align}
\kappa_{3\, jk}^{LL}  &= 1 - \left(\frac{ |y^{LL}_{3\, 3\ell}|^2 +|y^{LL}_{3\, 2\ell}|^2 }{16\pi^2}\right) \approx 0.996,
\end{align}
where we recall each LQ only couples to one lepton flavor, $\ell$, at a time. 
We multiply $y_{3\, jn}^{LL,\overline{\rm MS}}$ with ${\bar\kappa_{3\, jn}^{LL}} $.
to convert to the `low' scheme, and these conversion factors and the corresponding low-energy couplings are given in Table~\ref{tab:LQU}. We evaluate the $\overline{\text{MS}}$ conversion factor at $\mu=M_{S_{3}}$.

\begin{table}[t!]
\centering
\def\arraystretch{1.3}
\begin{tabular}{|c|c|c|c|c|}
\hline 
Benchmark & $S_{3\ell}$ & $(y^{LL,\overline{\text{MS}}}_{3\,2\ell},y^{LL,\overline{\text{MS}}}_{3\,3\ell})$ & 
$(\bar\kappa_{3\, 2\ell}^{LL} \kappa_{3\, 2\ell}^{LL},\bar\kappa_{3\, 3\ell}^{LL} \kappa_{3\, 3\ell}^{LL})$ & $(y^{LL,\text{low}}_{3\,2\ell},y^{LL,\text{low}}_{3\,3\ell})$ \\
\hline 
A & $S_{3\, e}$ &$(0.760, 0.189)$ & $(1.007,1.008)$ &$(0.765, 0.191)$\\
\cline{2-5}
 & $S_{3\, \mu}$ &$(0.191, 0.759)$ & $(1.005, 1.006)$ &$(0.192, 0.764)$\\
\cline{2-5}
 & $S_{3\, \tau}$ &$(0.639, -0.452)$ & $(1.006, 1.011)$ &$(0.643, -0.457)$\\
 \hline 
 B & $S_{3\, e}$ &$(0.189, 0.760)$ & $(1.007,1.007)$ &$(0.190, 0.765)$\\
\cline{2-5}
 & $S_{3\, \mu}$ &$(0.759, 0.191)$ & $(1.007, 1.008)$ &$(0.764,0.193 )$\\
\cline{2-5}
 & $S_{3\, \tau}$ &$(0.639, -0.452)$ & $(1.006, 1.011)$ &$(0.643, -0.457)$\\
 \hline 
\end{tabular}
\caption{Benchmark points from Ref.~\cite{Fedele:2023rxb}, converted to the low-scheme utilising our formalism.}
\label{tab:LQU}
\end{table}

Although $\bar\kappa_{3\,jk}^{LL}$ deviates from 1 more than $\kappa_{3\,jk}^{LL}$, the effects in Table~\ref{tab:LQU} are still small, because the fixed-point values of the couplings are moderate in size. In general, however, the first term in the square bracket in \eq{eq:kb} can give a large contribution, if the tree-level coupling is small.
\section{{Conclusions}}\label{sec:summary}
Leptoquark (LQ) phenomenology involves the combination of constraints from 
low and high energy. The former are inferred from solutions to the flavor anomalies, while the latter are found from LHC bounds on production cross sections. Stronger collider bounds, pushing the LQ masses to higher values, imply 
larger couplings in the LQ lagrangian in \eq{lag} to explain anomalous low-energy data. Sizable couplings, however, lead to potentially large loop corrections with LQs. We have studied
such corrections for the popular $S_1$, $R_2$, and $S_3$ LQ scenarios with the following results:

\textbf{(1)}~~The radiative corrections have a particularly simple structure,
as they can be completely absorbed into a renormalization of the LQ coupling.
This allows us to define two renormalization schemes designed for low-energy and 
high-energy observables. With the choice of our `low' scheme all considered corrections are absorbed into the coupling $y_{n\, jk}^{XY, \mathrm{low}}$, so that the tree and one-loop amplitudes for the studied flavor-changing processes 
coincide. The same feature holds for the LQ decay amplitudes in the `high' scheme. The combination of low-energy and 
high-energy data only involves the conversion factors 
$\kappa_{n, jk}^{XY}\equiv y_{n\, jk}^{XY, \mathrm{high}}/y_{n\, jk}^{XY, \mathrm{low}}$.  While this method captures all radiative corrections to 
both the low-energy process and the LQ decay amplitude, a full one-loop calculation 
of the production cross sections involves additional diagrams with process-dependent corrections and the use of $y_{n\, jk}^{XY, \mathrm{high}}$ in the tree result only captures a universal subset, but permit simulations with tree-level event generators.

\textbf{(2)} In scenarios with only one LQ species the $\kappa_{n, jk}^{XY}$
have a particularly simple structure, originating solely from self-energies,  
and are always proportional to the tree-level coupling. In these scenarios 
one always finds $\kappa_{n, jk}^{XY} < 1$, implying that collider constraints are weakened.
Only scenarios with both $S_1$ and $R_2$ involve vertex corrections, which moreover involve different couplings compared to tree level and can therefore be enhanced if the tree coupling is small. 

\textbf{(3)}~~Our calculations resulted in unexpectedly small loop functions, so that the  $\kappa_{n, jk}^{XY}$ are close to 1 for 
$y_{n\, jk}^{XY}={\cal O} (1)$. The LQ
scenarios explaining the $b\to c\tau\bar\nu$ and $b\to s\ell^+\ell^-$ anomalies
involve two couplings. In the $S_1$ and $S_3$ scenarios  ${\cal O} (1)$ couplings are compatible with LQ masses satisfying collider bounds and the $\kappa_{n, jk}^{XY}$'s can be neglected. However, for hierarchical choices of these couplings, with one coupling much larger than the other, $ \kappa_{n, jk}^{XY}$ can be substantially smaller than 1. 
 
\textbf{(4)}~~The smallness of the loop functions implies that perturbation theory works for couplings larger than 5. This feature opens up the parameter spaces, because 
low-energy data can be explained with large LQ masses and furthermore 
$\kappa_{n, jk}^{XY}<1$ weakens the collider bounds. 

\textbf{(5)}~~Like the vertex corrections mentioned in item (2) 
also flavor-changing fermion self-energies need not involve the tree-level coupling. While these self-energies drop out from the $\kappa_{n, jk}^{XY}$'s,
they contribute to the scheme conversion from $\overline{\rm MS}$ to 
either `low' or `high' scheme. The corresponding scheme transformation
factors $\bar\kappa_{n, jk}^{XY}$ are larger than the $\kappa_{n, jk}^{XY}$'s,
but can still be omitted for the choices of the infrared fixed-point solutions 
for $y_{3\, jk}^{LL,\overline{\mathrm{MS}}}$ found in 
Ref.~\cite{Fedele:2023rxb}.

\textbf{(6)}~~Irrespective of any low-energy data, collider searches  
first probe the parameter region with smallest masses and largest couplings. 
These searches will constrain $y_{n\, jk}^{XY, \mathrm{high}}$ and if these bounds 
are used in other schemes the couplings must be converted properly, e.g.\ the 
$\overline{\rm MS} $ values are found by dividing $y_{n\, jk}^{XY, \mathrm{high}}$ with
$\bar\kappa_{n, jk}^{XY}\kappa_{n, jk}^{XY}$.

\acknowledgments 
UN acknowledges the hospitality of the Fermilab Theory Group, where part of the presented work was completed.
This research was supported by Deutsche Forschungsgemeinschaft (DFG, German Research Foundation)  within the Collaborative Research Center \emph{Particle Physics Phenomenology after the Higgs Discovery (P3H)} (project no.~396021762 – TRR 257). This manuscript has been authored in part by Fermi Research Alliance, LLC under Contract No. DE-AC02-07CH11359 with the U.S. Department of Energy, Office of Science, Office of High Energy Physics. The work of I.B. was performed in part at the Aspen Center for Physics, supported by a grant from the Alfred P. Sloan Foundation (G-2024-22395).

\appendix

\section{Wilson coefficient evolution}\label{app:A}

In Ref.~\cite{Iguro:2022yzr}, the fit of the WCs to $b\to c\tau\nu$ is done at the scale $\mu_b=4.8$ GeV, which must then be evolved to the scale $m_\text{LQ}$ to be matched onto the full theory. The QCD renormalization-group equations~(RGEs)~\cite{Jenkins:2013wua,Alonso:2013hga,Gonzalez-Alonso:2017iyc} (first matrix) and the LQ-charge independent QCD one-loop matching~\cite{Aebischer:2018acj} (second matrix) give the following relation for $M_\text{LQ}=2$ TeV, 
\begin{align}
    \begin{pmatrix}
        C_{V_L}(\mu_b)\\
         C_{V_R}(\mu_b)\\
          C_{S_L}(\mu_b)\\
           C_{S_R}(\mu_b)\\
            C_{T}(\mu_b)
    \end{pmatrix} \nonumber
    &\approx 
    \begin{pmatrix}
        1 & 0 &0 &0 &0\\
        0 & 1 &0 &0 &0\\
        0 & 0 &1.82 &0 &-0.35\\
        0 & 0 &0 &1.82 &0\\
        0 & 0 &-0.004 &0&0.83\\
    \end{pmatrix}
    \begin{pmatrix}
        1.12 & 0 &0 &0 &0\\
        0 & 1.07 &0 &0 &0\\
        0 & 0 &{1.05} &0 &0\\
        0 & 0 &0 &{1.10} &0\\
        0 & 0 &0&0&1.07\\
    \end{pmatrix}
     \begin{pmatrix}
        C_{V_L}(M_\text{LQ})\\
         C_{V_R}(M_\text{LQ})\\
          C_{S_L}(M_\text{LQ})\\
           C_{S_R}(M_\text{LQ})\\
            C_{T}(M_\text{LQ})
    \end{pmatrix},\\
   & \approx \begin{pmatrix}
        1.12 & 0&0&0&0\\
        0 & 1.07&0&0&0\\
        0 & 0&1.91&0&-0.38\\
        0 & 0&0&2.00&0\\
        0 & 0&0&0&0.89\\
    \end{pmatrix}
      \begin{pmatrix}
        C_{V_L}(M_\text{LQ})\\
         C_{V_R}(M_\text{LQ})\\
          C_{S_L}(M_\text{LQ})\\
           C_{S_R}(M_\text{LQ})\\
            C_{T}(M_\text{LQ})
    \end{pmatrix},
\end{align}
from which we obtain $C_{S_L}(\mu_b)\approx -8.9 C_T(\mu_b)$ for $S_1$ and $C_{S_L}(\mu_b)\approx 8.4 C_T(\mu_b)$ for $R_2$.

\section{A note on constraints for flavour anomaly models}\label{app:B}

We briefly comment on constraints relevant for selecting the benchmark points in the main text.
\subsection{\texorpdfstring{Constraints from $b\to s\nu\bar{\nu}$}{constraints bsnuu}}
The contribution of BSM to $b\to s\nu\overline{\nu}$ transitions may be parameterized by the following, in the absence of significant right-handed vector currents~\cite{Buras:2014fpa} 
\begin{align}
    \mathcal{L}^{\nu\bar{\nu}} \supset \frac{4G_F}{\sqrt{2}} V_{tb} V_{ts}^* C_{\nu\bar{\nu}}^\ell \left[\frac{\alpha_{\text{em}}}{4\pi}(\bar{s}\gamma_\mu P_L b)(\bar{\nu}_\ell \gamma^\mu(1-\gamma_5)\nu_\ell) \right] +\text{h.c.}.
\end{align}
As collider experiments do not distinguish between neutrino flavors, the sum over all flavors appears in the ratio of branching fractions and its SM prediction
\begin{align}
    R_{K^{(*)}}^{\nu\bar{\nu}} = \frac{\text{Br}^\text{exp}(B\to K^{(*)}\nu\bar{\nu})}{\text{Br}^\text{SM}(B\to K^{(*)}\nu\bar{\nu})}= \frac{(C_{\nu\bar{\nu}}^\text{SM}+C_{\nu\bar{\nu}}^e)^2+(C_{\nu\bar{\nu}}^\text{SM}+C_{\nu\bar{\nu}}^\mu)^2+(C_{\nu\bar{\nu}}^\text{SM}+C_{\nu\bar{\nu}}^\tau)^2}{3(C_{\nu\bar{\nu}}^\text{SM})^2},
\end{align}
where $C_{\nu\bar{\nu}}^\text{SM}\approx -6.35$. The experimental limits from the Belle collaboration read    $R_{K}^{\nu\bar{\nu}}<3.9$ and   $ R_{K^{*}}^{\nu\bar{\nu}}<2.7$ at 90\% C.L.\cite{Belle:2017oht}. The recent Belle-II measurement of Br($B^+\to K^++\text{inv.})=(2.4\pm 0.7)\times 10^{-5}$\cite{Belle-II:2023esi}, assuming three-body decay with massless neutrinos, may be contrasted with the  SM prediction for this branching ratio, $\text{Br}^\text{SM}(B^+\to K^+\nu\bar{\nu})= (2.81\pm 0.15)\times 10^{-3} \times V_{tb}V_{ts}^*$~\cite{Parrott:2022zte}. This results in $R_{K}^{\nu\bar{\nu}}\approx 5$, which exceeds the previous constraint, and corresponds to a $\approx 2.8 \sigma$ deviation from the SM expectation. In the absence of considerable right-handed vector currents, $R_{K^{*}}^{\nu\bar{\nu}}=R_{K}^{\nu\bar{\nu}}$, and the Belle II measurement of Ref.~\cite{Belle-II:2023esi} is in conflict with the experimental bound on $R_{K^{*}}^{\nu\bar{\nu}}$ \cite{Bause:2023mfe} and therefore only the latter will be used as a constraint. 

For the considered scenarios in the main text we consider tree-level contributions from the $S_1$ singlet
\begin{align}
    C_{\nu\bar{\nu}}^\tau =\frac{\pi }{\sqrt{2} G_F V_{tb} V_{ts}^* \alpha_\text{em}}\frac{y_{1\; 33}^{LL} y_{1\; 23}^{LL\, *}}{M_{S_1}^2},
\end{align}
and $C_{\nu\bar{\nu}}^e=C_{\nu\bar{\nu}}^\mu=0$, and for the lepton-flavor universal triplet LQ $S_3^\ell$
\begin{align}
    C_{\nu\bar{\nu}}^\ell =\frac{\pi }{\sqrt{2} G_F V_{tb} V_{ts}^* \alpha_\text{em}}\frac{y_{3\; 3\ell}^{LL} y_{3\; 2\ell}^{LL\, *}}{M_{S_1}^2}.
\end{align}
 We find the benchmark points considered in the main text satisfy the constraint from $R_{K^{*}}^{\nu\bar{\nu}}$ for both LQ models. Reconciling the experimental results for $R_{K^{*}}^{\nu\bar{\nu}}$ and $R_{K}^{\nu\bar{\nu}}$ requires new physics couplings to right-handed $b$ and $s$ quarks (see e.g. Ref.~\cite{Bause:2023mfe} for discussion), which  $S_{3\,\ell}$ and $S_1$ cannot provide at tree-level.

\subsection{\texorpdfstring{Constraints from $Z\to \tau\overline{\tau}$}{constraints Ztautau}}

For calculating the corrections to  $Z$ coupling to leptons, we follow the procedure of Ref.~\cite{Arnan:2019olv}. To parametrize these effects, we consider the matrix element of the decay of a  $Z$ boson into a SM fermion-antifermion pair  
$(f_i,\bar f_j)$,
\begin{align}
  \mathcal{M} (Z\to f_i\bar f_j) = \frac{g}{\cos\theta_W}  
    \bar{u}_i \gamma^\mu \left[ g^{ij}_{f_L} P_L + g^{ij}_{f_R} P_R\right] v_j \epsilon_\mu^Z \, ,
\end{align}
where $g$ is the SU$(2)_L$ gauge coupling, $u$ and $v$ are spinors, $\epsilon_\mu^Z$ is the 
$Z$ polarization vector,and 
\begin{align}
    g_{f_{L(R)}}^{ij}=  g^{\text{SM}}_{f_{L(R)}} \delta^{ij}+ \delta g^{ij}_{f_{L(R)}}.\label{definedcoup}
\end{align}
At tree level, the SM effective couplings are given by $g^{0}_{f_{L}}= T_3^f- Q^f \sin^2 \theta_W$ and $ g^{0}_{f_{R}}=- Q^f \sin^2 \theta_W$,  where $Q^f$ is the electric charge of the fermion $f$, and $T^f_3$ is its third component of weak isospin.  At higher loop order in the SM, these couplings are modified by factors $\rho_f=1.00937$ and $\sin^2 \theta_\text{eff}=0.231533$~\cite{ParticleDataGroup:2020ssz}, 
\begin{align}
g^\text{SM}_{f_{L}}= \sqrt{\rho_f} \, (T_3^f- Q^f \sin^2 \theta_\text{eff}), \hspace{2cm} g^\text{SM}_{f_{R}}=- \sqrt{\rho_f} \, Q^f \sin^2 \theta_\text{eff}. \label{effectivecouplingsZ}
\end{align}
 Focusing on the effective coupling to charged leptons, $f=\ell$, and noting that $g_{\ell,{V(A)}}^{ij}=g_{\ell_L}^{ij}\pm g_{\ell_R}^{ij}$, we constrain the combination $g_{\ell,{V(A)}}^{ii}/g^\text{SM}_{\ell}$ for the lepton-flavor-diagonal couplings.

For the doublet LQ $R_2$ as a solution to anomalies in $b\to c\tau\nu$, dominant constraints arise from loop-order corrections to $Z \to \tau\overline{\tau}$ couplings~\cite{Becirevic:2018uab}. The leading-order contributions to perturbing the effective couplings are given by $\delta g_{\tau_R}=g_{\ell_R}^{33}$  and $\delta g_{\tau_L}=g_{\ell_L}^{33}$. The loop corrections have been calculated as in \cite{Becirevic:2018uab},
\begin{align}
    \delta g_{\tau_R}&= N_c |y^{LR}_{2\,33}|^2\Bigg[ \frac{x_t}{32\pi^2}(1+\log x_t)\\&\;\;\;\;+\frac{x_Z}{144\pi^2}\left[-\left(\sin^2\theta_W -\frac{3}{2}\right) (\log x_z+i\pi)+\left( -\frac{1}{4}+\frac{2}{3}\sin^2\theta_W\right)\right]\Bigg],\nonumber\\
\delta g_{\tau_L}&= N_c |y^{RL}_{2\,23}|^2 \frac{x_z}{72\pi^2}\left[ \sin^2\theta_W\left(\log x_z+i\pi+\frac{1}{12}\right)-\frac{1}{8}\right],
\end{align}
taking $V_{tb}\approx 1$, $N_c=3$ is the number of colours, $\sin^2\theta_W$ is taken to be $\sin^2 \theta_\text{eff}$,  $x_t= m_t^2/m_{R_2}^2$ and $x_Z= m_Z^2/m_{R_2}^2$. These effective couplings given above are constrained by LEP measurements of the $Z$
decay widths and other electroweak observables~\cite{ALEPH:2005ab}. Specifically, for the effective coupling to the tau, the strongest constraints come from the axial-vector coupling, $\mathrm{Re}(g_{\tau_{A}})/g_{A}^{\rm SM}=1.00154 \pm0.00128$ which we require to be within two-sigma of the central value~\cite{Crivellin:2020mjs}. The benchmark point examined in Section~\ref{scalardoublet} satisfies this constraint.

\section{Extending to other scalar leptoquark models}\label{app:C}

In this work, we highlight the three scalar leptoquarks $S_1, R_2$ and $S_3$ relevant for flavour-anomaly model building. Here we extend the calculations in Section~\ref{sec:setup} to models featuring the two other scalar LQs: $\tilde{S}_1\sim ({\bar{{3}}}, {{1}}, 4/3)$ and $\tilde{R}_2\sim ({{3}}, {2}, 1/6)$. These extend the Lagrangian in Eq.~\eqref{lag} to include~\cite{Dorsner:2016wpm}
\begin{equation}
    \mathcal{L}_{\text{LQ}}\supset \sum_{j, k =1}^3 \sum_{a, b=1}^2 \left[\tilde{y}_{1, ij}^{RR} \overline{d^c_R}^i \tilde{S}_1 e_R^j - \tilde{y}_{2\, ij}^{RL} \overline{d^i}_R \tilde{R}_2^a \epsilon^{ab} L_L^{j, b} + \text{h.c..}\right]
\end{equation}
With the addition of these scalars, the only one-loop corrections to LQ vertices are from their respective self-energy corrections. This extends the list of results in Eq.~\eqref{eq:correction} to include
\begin{align}
      \tilde{\kappa}_{2\, jk}^{RL}  &=1 - \frac{ \sum_{l,n=1}^3   |\tilde{y}_{2\ ln}^{RL}|^2   }{32\pi^2}, \; \\
      \tilde{\kappa}_{1\, jk}^{RR}  &=1 - \frac{ \sum_{l,n=1}^3   |\tilde{y}_{1\ ln}^{RR}|^2}{32\pi^2}, \; 
\end{align}
which may be integrated into the correction formula akin to those for the LQs discussed in the main text. 

\bibliographystyle{JHEP}
\bibliography{bibl}

\end{document}